\documentclass[runningheads]{llncs}
\usepackage{eccv}
\usepackage{eccvabbrv}
\usepackage{booktabs}
\usepackage{colortbl}
\usepackage{makecell}
\usepackage{marvosym}
\usepackage{float}
\usepackage{graphicx}
\usepackage{booktabs}
\usepackage{multirow}
\usepackage{placeins}
\usepackage{afterpage}
\usepackage[accsupp]{axessibility} 
\usepackage{hyperref}
\usepackage{orcidlink}
\usepackage{color}

\newcommand{\ieno}{\textit{i.e.}}
\newcommand{\egno}{\textit{e.g.}}
\begin{document}
\title{MoE-DiffIR: Task-customized Diffusion Priors for Universal Compressed Image Restoration}
\titlerunning{MoE-DiffIR}
\author{Yulin Ren\inst{1}\orcidlink{0009-0006-4815-7973} \and
Xin Li\inst{1}\orcidlink{0000-0002-6352-6523}\textsuperscript{~(\Letter)} \and
Bingchen Li\inst{1}\orcidlink{0009-0001-9990-7790} \and
Xingrui Wang\inst{1}\orcidlink{0009-0009-3083-0658} \and
Mengxi Guo\inst{2}\orcidlink{0009-0007-9490-6661} \and
Shijie Zhao\inst{2}\orcidlink{0000-0002-8466-8061} \and
Li Zhang \inst{2}\orcidlink{0000-0002-3463-9211} \and
Zhibo Chen\inst{1}\orcidlink{0000-0002-8525-5066}\textsuperscript{~(\Letter)}}
\authorrunning{Y.~Ren et al.}
\institute{
University of Science and Technology of China, Hefei, Anhui, China \\
\and
Bytedance Inc., Beijing, China \\
\email{\{renyulin, lbc31415926, wxrui\_18264819595\}@mail.ustc.edu.cn} \\
\email{\{xin.li, chenzhibo\}@ustc.edu.cn}, \email{nicolasguo@pku.edu.cn} \\
\email{\{zhaoshijie.0526, lizhang.idm\}@bytedance.com}
}

\maketitle
\renewcommand{\thefootnote}{}
\footnotetext{\textsuperscript{~(\Letter)}  Corresponding authors.}

\renewcommand{\thefootnote}{\arabic{footnote}}
\setcounter{footnote}{0}

\begin{abstract}
We present MoE-DiffIR, an innovative universal compressed image restoration (CIR) method with task-customized diffusion priors. This intends to handle two pivotal challenges in the existing CIR methods: (i) lacking adaptability and universality for different image codecs, \eg, JPEG and WebP; (ii) poor texture generation capability, particularly at low bitrates. Specifically, our MoE-DiffIR develops the powerful mixture-of-experts (MoE) prompt module, where some basic prompts cooperate to excavate the task-customized diffusion priors from Stable Diffusion (SD) for each compression task. Moreover, the degradation-aware routing mechanism is proposed to enable the flexible assignment of basic prompts. To activate and reuse the cross-modality generation prior of SD, we design the visual-to-text adapter for MoE-DiffIR, which aims to adapt the embedding of low-quality images from the visual domain to the textual domain as the textual guidance for SD, enabling more consistent and reasonable texture generation. We also construct one comprehensive benchmark dataset for 
universal CIR, covering 21 types of degradations from 7 popular traditional and learned codecs. Extensive experiments on universal CIR have demonstrated the excellent robustness and texture restoration capability of our proposed MoE-DiffIR. The project can be found at~\textcolor{magenta}{~\url{https://renyulin-f.github.io/MoE-DiffIR.github.io/}}.

\end{abstract}

\section{Introduction}
\label{sec:intro}

\begin{figure}[t]
	\centering
	\includegraphics[width=1\linewidth]{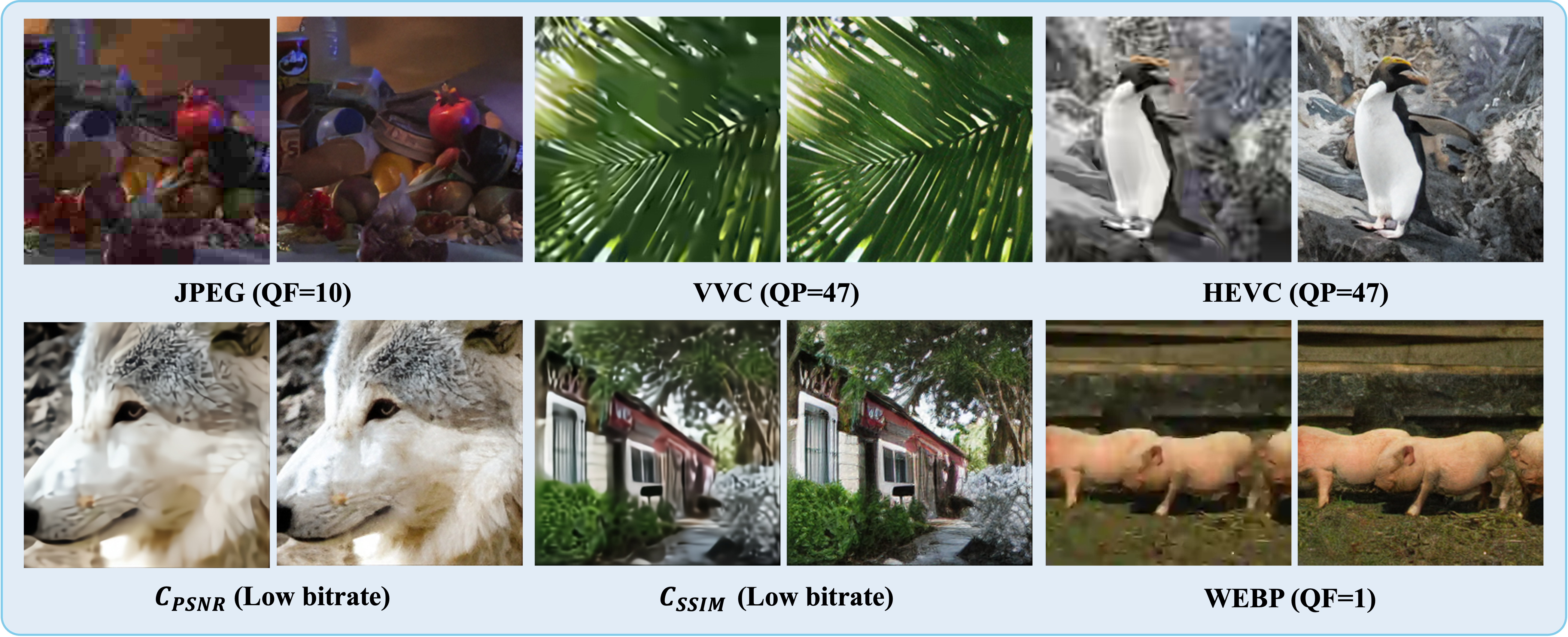}
	\caption{Visualization of restored compressed images with our MoE-DiffIR on various image codecs and coding modes. Our method can restore diverse compressed images at low bitrates through a single network while possessing high texture generation capability.}
	\label{fig:Current challenges}
\end{figure}

Image compression has emerged as a ubiquitous and indispensable technique in human life and industrial applications, aiming to reduce the costs of image transmission and storage. Existing image codecs can be roughly divided into two categories: (i) traditional image codecs~\cite{bross2021overview-VVC,sullivan2012overview-HEVC,christopoulos2000jpeg2000,wallace1991jpeg,li2021task-code-task}, which are designed based on elaborate pre-defined transform, and coding modes, \egno, JPEG~\cite{wallace1991jpeg}, BPG~\cite{yee2017medical-BPG}, and WebP~\cite{ginesu2012objective-WEBP} etc; (ii) learned end-to-end image codec~\cite{he2022elic-compression_new2,wu2021learned,agustsson2023multi_compression_new4}, where rate-distortion optimization is achieved with learnable non-linear transform, soft quantization, entropy coding, and other techniques. Despite the substantial success, the compressed images inevitably encounter severe compression artifacts, such as blur, color shift, and block artifacts at low bitrate, which brings an unpleasing visual experience as shown in Fig.~\ref{fig:Current challenges}. 

To remove the complicated compression artifacts, the Compressed Image Restoration (CIR) task has been extensively investigated by a series of pioneering studies~\cite{foi2007pointwise-model-based1,Filter-model-based2,yang2022aim}, focusing on the design of the restoration network. Based on advanced Convolution Neural Networks (CNN)~\cite{krizhevsky2012imagenet-CNN} and Transformer~\cite{vaswani2017attention} architecture, some works~\cite{dong2015compression-Learning-based1,li2020learning-disentangle,jiang2021towards-FBCNN,ehrlich2020quantization-QGAC,chen2023hat,yang2024ntire-cir,li2024sed} achieved excellent objective performance (\egno, PSNR, SSIM) on JPEG artifact removal. However, as shown in Fig.~\ref{fig:Current challenges}, these works overlooked two essential challenges in the current CIR task: (i) numerous image codecs and coding modes, leading to diverse compression artifacts. For instance, at low bitrate, the JPEG codec tends to produce blocking artifacts, whereas the learned codecs, \egno, $C_{PSNR}$~\cite{cheng2020learned} is susceptible to more blur artifacts. This raises an urgent requirement for an all-in-one/universal CIR method; (ii) unsatisfied texture recovery due to the lack of generation priors in low-quality images and CIR models. 

To address the above challenges, we aim to achieve an all-in-one CIR model by excavating diffusion priors from Stable Diffusion (SD)~\cite{wang2023exploiting-stablesr,lin2023diffbir,yang2023pixel-PASD,jiang2023autodir,ai2023multimodal-mperceiver,chung2023prompt-P2L}. Notably, existing works have shown the superior applicability of stable diffusion in image restoration, \egno, StableSR~\cite{wang2023exploiting-stablesr}, and DiffBIR~\cite{lin2023diffbir}, which reuse the generation priors of diffusion models for a specific task by introducing the modulation module, like ControlNet~\cite{zhang2023adding-Controlnet}, feature adapter~\cite{ye2023ip-ipadapter,mou2023t2i,zhao2024uni-adapter}. Nonetheless, the above approaches are inadequate for effectively modulating diffusion models for multiple CIR tasks with shared modulation parameters. Recently, prompt learning has demonstrated its potential and efficiency for universal image restoration framework~\cite{potlapalli2023promptir,li2023prompt-PIP,ma2023prores,luo2023controlling-DACLIP,ai2023multimodal-mperceiver,gou2023exploiting-all-in-one}. Inspired by this, we explore how to utilize prompt learning to simultaneously excavate diffusion priors within Stable Diffusion for multiple CIR tasks.

In this work, we present MoE-DiffIR, the first all-in-one diffusion-based framework for universal compressed image restoration with prompt learning. Particularly, the various CIR tasks usually own distinct degradation forms due to different image codecs/modes. This entails the requirement of the task-customized diffusion priors for each CIR task from Stable Diffusion. To this end, we propose the advanced Mixture-of-Experts (MoE) Prompt module, which takes advantage of MoE~\cite{shazeer2017outrageously-MoE2,masoudnia2014mixture-MoE1,zhou2022mixture-MoE-latest} to enable dynamic prompt learning for multiple CIR tasks with fixed few prompts. Concretely, we set a series of basic prompts as degradation experts, and design the degradation-aware router to customize the modulation scheme for each task by adaptively selecting top $K$ prompts. In contrast to single prompt or multiple weighted prompts in ~\cite{ma2023prores,potlapalli2023promptir,ai2023multimodal-mperceiver,luo2023controlling-DACLIP}, our MoE-Prompt enables each prompt to perceive different degradations and improve the parameter reuse.

It is noteworthy that Stable Diffusion possesses a rich text-to-image generation prior, which is usually overlooked by existing IR works~\cite{lin2023diffbir,wang2023exploiting-stablesr}. 
To activate and reuse these cross-modality priors, we introduce the visual-to-text adapter. Particularly, the CLIP visual encoder is exploited to extract the visual embedding from low-quality images, and the visual-to-text adapter is responsible for transforming the visual embedding into corresponding textual embedding for the guidance of Stable Diffusion. Considering that the low-quality image might damage the extracted visual embedding, we utilize several transform layers as the quality enhancer before the CLIP visual encoder. To validate the effectiveness of our MoE-DiffIR, we construct the first benchmark for the universal CIR task by collecting 7 commonly used image codecs, including 4 traditional codecs and 3 learnable codecs, each with three levels of compression, resulting in 21 types of degradations. Extensive experiments on the universal CIR task have shown the superiority of our MoE-DiffIR in terms of improving perceptual quality and enhancing the robustness for various compression artifacts.

The main contributions of this paper are as follows:
\begin{itemize}
    \item We propose the first all-in-one diffusion-based method for universal compressed image restoration (CIR) by extracting the task-customized diffusion priors from Stable Diffusion for each CIR task.  
    \item Based on the Mixture-of-Experts (MoE), we propose the MoE-Prompt module to enable each prompt expert to perceive the different degradation and cooperate to extract task-customized diffusion priors. Moreover, we active and reuse the cross-modality generation priors with our proposed Visual-to-Text adapter, which further uncovers the potential of stable diffusion. 
    \item We construct the first dataset benchmark for the CIR tasks, consisting of 7 typical traditional and learned image codecs/modes, each with 3 compression levels, resulting in 21 types of degradation tasks. 
    \item Extensive experiments on 21 CIR tasks have validated the effectiveness of our proposed MoE-DiffIR in improving the perceptual quality and the excellent robustness for unseen compression artifacts. 
\end{itemize}

\section{Related Work}

\subsection{Compressed Image Restoration}
Compressed image restoration (CIR) aims to restore compressed images generated by different codecs at varying bitrates. Existing CIR methods typically employ CNN-based~\cite{li2020multi-vvc,dong2015compression-Learning-based1,ehrlich2020quantization-QGAC,jiang2021towards-FBCNN} or Transformer-based approaches~\cite{liang2021swinir,wang2022jpeg-contrastive,chen2023hat,li2024promptcir}. QGAC~\cite{ehrlich2020quantization-QGAC} and FBCNN~\cite{jiang2021towards-FBCNN} are typical CNN-based methods that predict quality factors of compressed images to achieve blind restoration of JPEG codecs. The work~\cite{wang2022jpeg-contrastive} proposes an unsupervised compression encoding representation learning method specifically for JPEG, improving generalization in the JPEG domain.
However, these methods primarily aim to enhance the objective quality of the restored images and have poor perceptual quality at extremely low compression bitrates. Additionally, they only target a specific compression codec like JPEG, lacking generality in practical applications.

\subsection{Diffusion-based Image Restoration}
The impressive generative capabilities of diffusion models hold potential for various visual domains, including low-level vision tasks~\cite{moser2024diffusion-survey-2,yang2023diffusion-survey-3,kawar2022denoising-ddrm,jin2024des3}. Diffusion-based image restoration (IR) methods can be divided into two categories~\cite{li2023diffusion-survey}: supervised IR methods~\cite{saharia2022image-SR3,luo2023image-IRSDE,luo2023refusion,li2022srdiff,xia2023diffir} and zero-shot IR methods~\cite{wang2022zero-ddnm,kawar2022denoising-ddrm,chung2022diffusion-dps,rout2024solving-PLSD,wang2023dr2}. Recently, some works~\cite{wang2023exploiting-stablesr,lin2023diffbir,gou2023exploiting-all-in-one,jiang2023autodir} have attempted to fine-tune pre-trained SD models to extract diffusion priors for real-world image restoration. The pioneering work in this area is StableSR~\cite{wang2023exploiting-stablesr}, which fine-tunes a pre-trained Stable Diffusion model with a time-aware encoder for image restoration in real-world scenes. Another method is DiffBIR~\cite{lin2023diffbir}, which combines SwinIR~\cite{liang2021swinir} to first perform coarse-level restoration of distorted images and then utilizes Stable Diffusion with ControlNet~\cite{zhang2023adding-Controlnet} for details refinement. PASD~\cite{yang2023pixel-PASD} attempts to employ pre-trained BLIP and ResNet models to extract high-level information from low-quality images to directly guide the Stable Diffusion restoration.

\subsection{Prompt Learning in Image restoration}
Recently, prompt learning has significantly influenced the fields of language and vision~\cite{nie2023pro-prompt-latest1,zhu2023prompt-prompt-latest2,jia2022visual-prompt-latest3}. Several studies have begun applying prompts to low-level tasks, with PromptIR~\cite{potlapalli2023promptir} being a notable example. This work extends Restormer~\cite{zamir2022restormer}, introducing a set of prompts to identify different distortions, and uses soft weights to manage these prompts for all-in-one image restoration. Another pioneering work is ProRes~\cite{ma2023prores}, which employs a singular image-like prompt to interact with various distortions. Additionally, PIP~\cite{li2023prompt-PIP} suggests a dual-prompt approach: one type for universal texture perception and another suite for different degradation types, similar to the weighting approach of PromptIR. In diffusion-based methods, DACLIP\cite{luo2023controlling-DACLIP} also incorporates multiple prompts with soft weight combinations at each time step, facilitating multi-task learning.

Unlike previous prompt-based methods, this paper leverages the concept of routers within the Mixture of Experts (MoE) framework, treating different prompts as experts for routing. It schedules combinations of prompts based on different distortion tasks. In this way, basic prompts can cooperate to fully excavate the task-customized diffusion priors for multiple CIR tasks.

\section{Proposed Method}
\label{sec:Methodology}
We propose a novel framework dubbed MoE-DiffIR for universal compressed image restoration. Firstly, we review the concept of Mixture-of-Experts and Stable Diffusion in Sec.~\ref{preliminary}. In order to fully excavate the task-customized diffusion priors from stable diffusion, we propose a mixture of experts prompt module illustrated in Sec.~\ref{MoE Prompts Methodology}. Meanwhile, we design the visual-to-text adapter for MoE-DiffIR in Sec.~\ref{Visual2Text} to generate more realistic and consistent texture. Additionally, we introduce the entire framework and fine-tuning process of MoE-DiffIR in Sec.~\ref{Overview fine-tuning process}. Finally, we present our proposed dataset benchmark for compressed image restoration tasks in Sec.~\ref{CIR dataset benchmark}.

\subsection{Preliminary}
\label{preliminary}
\paragraph{Mixture-of-Experts:}
The Mixture of Experts (MoE) model is an effective method for increasing the capabilities~\cite{masoudnia2014mixture-MoE1,shazeer2017outrageously-MoE2,luo2023image-MoE3} of the models, and it is frequently employed in various scaling-up tasks. In MoE, routers select and activate different experts based on the input tokens using various routing mechanisms~\cite{masoudnia2014mixture-MoE1,puigcerver2023sparse-soft-MoE,zhou2022mixture-MoE-latest}. A particularly typical example is the use of Sparsely Gated MoE~\cite{shazeer2017outrageously-MoE2}  where the output $y$ of MoE layer could be described as:
\begin{equation}
    y = \sum_{i=1}^{n} G(x)_i E_i(x)
\end{equation}
Here, $G_{x}$ and $E_{i}$ denote the output of router and $i$-th expert, respectively. In this work, we draw inspiration from the routing concept in MoE framework to combine basic prompts, which enables the prompts to cooperate together and fully excavate the task-customized diffusion priors for universal compression tasks.

\paragraph{Stable Diffusion:}
Stable diffusion conducts diffusion process in latent space, where a VAE encoder is used to compress input image into latent variable $z_{0}$. Then the model predict added noise to noisy latent $z_{t}$ with a unet network. The optimization function could be written as follows:
\begin{equation}
    \mathcal{L}_{\text{SD}} = \mathbb{E}_{\epsilon \sim \mathcal{N}(0,1)}\left[\| \epsilon - \epsilon(z_t, t) \right\|_{2}^2]
\end{equation}
Where $t$ denotes the time step and $\epsilon$ denotes the noisy map to be estimated.

\begin{figure}[htp]
	\centering
	\includegraphics[width=1\linewidth]{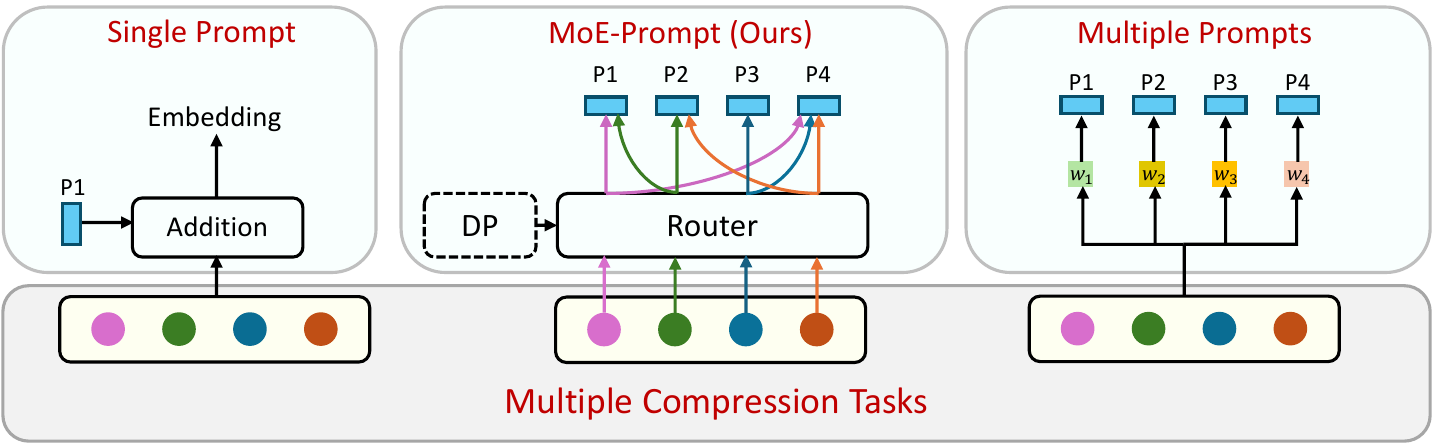}
	\caption{Comparison of different prompt interaction methods. Here we mainly categorize them into three types: (a) Single Prompt~\cite{ma2023prores}, (c) Multiple Prompts~\cite{li2023prompt-PIP,luo2023controlling-DACLIP,ai2023multimodal-mperceiver}, (b) MoE-Prompt (Ours). We use Mixture of Experts routing methods to select different combinations of prompts for various compression tasks. In (b), DP stands for Degradation Prior which is obtained from LQ images through pre-trained CLIP encoder of DACLIP.}
	\label{fig:Promps-compare}
\end{figure}

\subsection{Mixture-of-Experts Prompt}
\label{MoE Prompts Methodology}
As depicted in Fig.~\ref{fig:Promps-compare}(b),  we propose the mixture-of-experts (MoE) prompt to excavate task-customized diffusion priors. Unlike previous prompt-based IR methods, MoE-Prompt is designed to maximize reusability and the representative capacity of each prompt expert for different tasks. Concretely, there are two commonly used prompt-based IR categories. The first category is the single prompt, as shown in Fig.~\ref{fig:Promps-compare}(a), where a single prompt (usually image-like) is used to perceive distortions from different tasks through simple addition. This method struggles to model multiple tasks effectively, particularly as the number of tasks increases. A single prompt makes it difficult to manage complex relationships between different tasks.

The second category involves the use of multiple prompts, as represented in Fig.~\ref{fig:Promps-compare}(c), in most works~\cite{ai2023multimodal-mperceiver,potlapalli2023promptir,luo2023controlling-DACLIP}. Specifically, these methods set a prompt pool and generate a set of weights: $w_{1},w_{2},...,w_{n}$, which are used to multiply the predefined prompts and fuse them with soft weighting. However, this method is susceptible to the ``mean feature'', \ieno, these prompts learn similar features, lacking the diversity and reducing the modulation capability of universal tasks (Please see Sec.~\ref{Appendix:MoE Prompts} in the \textbf{Appendix}). The reason is due to the lack of one mechanism to enable these prompts to learn distinct degradation/task knowledge.

Therefore, the core principle of our MoE-Prompt method is to treat each prompt as an expert, allowing for the adaptive selection and scheduling of the necessary prompt cooperation for different distortion tasks through a router. This enables prompts to better cooperate and be reused for extracting task-customized diffusion priors. As depicted in Fig.~\ref{fig:Promps-compare}(b), it is necessary to provide distortion-related information to the router. Considering that DA-CLIP~\cite{luo2023controlling-DACLIP} is trained on large-scale distortion tasks and has demonstrated robustness to out-of-domain data, we use the pre-trained CLIP encoder from DACLIP to extract the degradation prior (DP) from low-quality images for various compression tasks. The obtained DP interacts with input features through a cross-attention mechanism and is then fed into the router. A more detailed diagram of this structure could be found in the Sec.~\ref{Appendix:MoE Prompts} of the \textbf{Appendix}. After that, the router adaptively selects a combination of prompts using a noisy Top-K function~\cite{shazeer2017outrageously-MoE2}, which is formalized as:
\begin{equation}
    G(x) = \text{Top-K}(\text{Softmax}(xW_g + \mathcal{N}(0, 1)\text{Softplus}(xW_{\text{noise}})))
\end{equation}
where $x$ represents the input features, $W_{g}$ is the weight matrix for global features, and $W_{noise}$ introduces stochasticity to the selection process, encouraging robustness and diversity in prompt selection. ``Softplus'' here is the smooth approximation to the ReLU function. Once $K$ prompts have been selected, they interact with the input feature through a form of matrix multiplication.
\begin{figure}[t]
	\centering
	\includegraphics[width=1\linewidth]{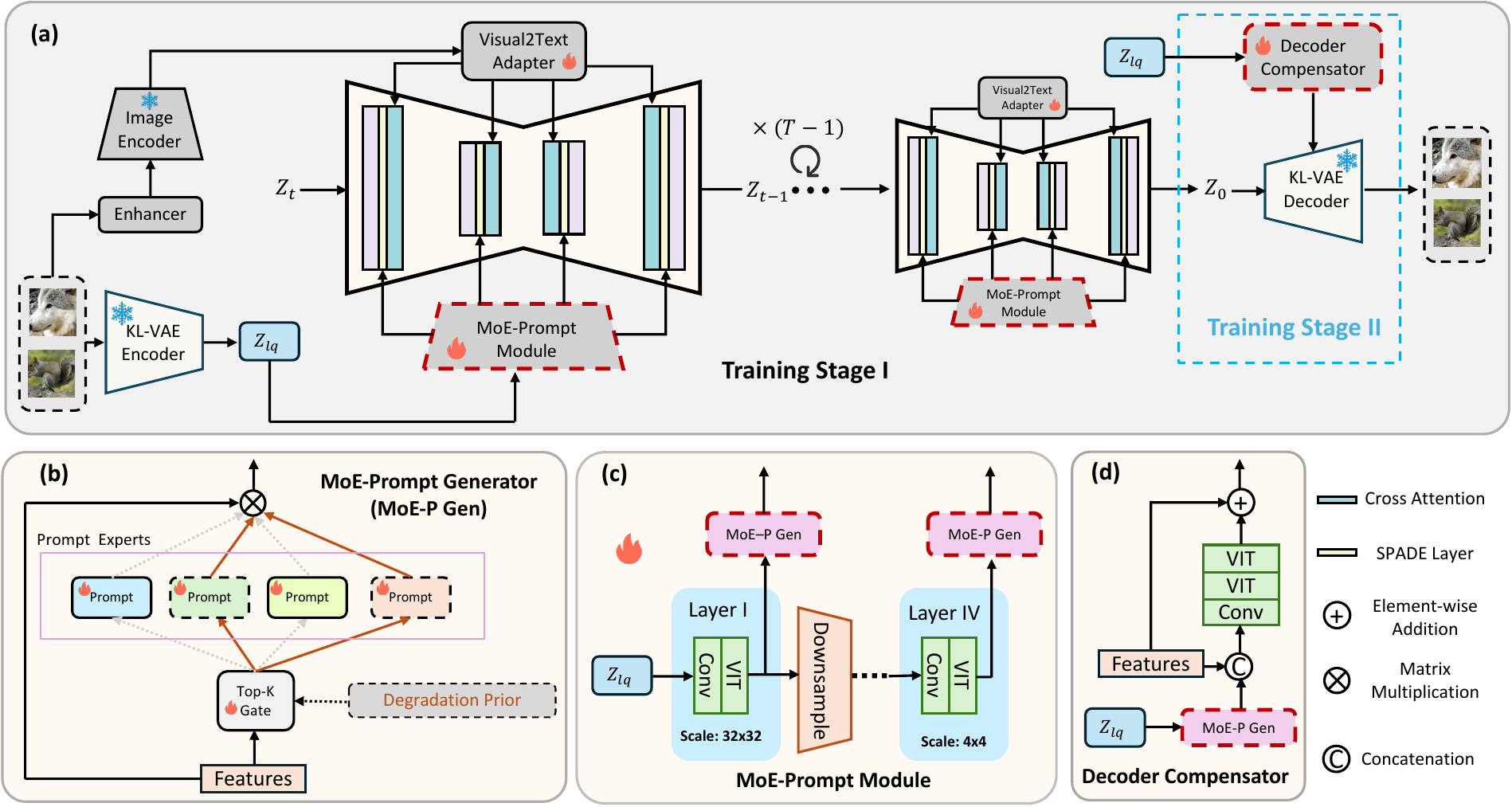}
	\caption{The framework of the proposed MoE-DiffIR enables dynamic prompt learning for multiple CIR tasks through (b) MoE-Prompt Generator, and introduces a visual-to-text adapter to generate more reasonable texture. In MoE-DiffIR: MoE-Prompt Module (c) aims to extract multi-scale features to interact with (b). Here (a) depicts the process of fine-tuning Stable Diffusion, which consists of two stages. Stage I: only the MoE-Prompt Module is pre-trained to excavate task-customized diffusion priors for each CIR task. Stage II: the (d) Decoder Compensator is fine-tuned for structural correction.}
	\label{fig:MoE-DiffIR}
\end{figure}

\subsection{Visual2Text Adapter}
\label{Visual2Text}
Stable Diffusion, trained on large-scale datasets~\cite{schuhmann2022laion-5b}, stores an abundance of text-to-image priors. However, these priors are often overlooked by some existing SD-based IR works.  For instance, StableSR~\cite{wang2023exploiting-stablesr} and DiffBIR~\cite{lin2023diffbir} configure the text condition input for the SD as an empty string. In order to activate and reutilize textual prior knowledge, we attempt to extract visual information from low-quality(LQ) images and transform it into the text embedding space. Indeed, there are some attempts to leverage pre-trained language models like BLIP for direct textual feature extraction from LQ images, such as PASD~\cite{yang2023pixel-PASD}. However, in the realm of compressed image restoration (CIR), especially at very low bit rates, the damage to the distorted images is severe. Extracting textual features from these images could potentially degrade the performance of the model.  Therefore, as shown in Fig.~\ref{fig:MoE-DiffIR}(a), we first enhance the LQ images using several transformer blocks and then employ CLIP’s image encoder to directly extract visual features. To better leverage the robust text-to-image capabilities of SD, we employ several MLP layers~\cite{gao2024clip-adapter} (referred to the Visual2Text Adapter) to translate visual information into the textual domain of SD. This approach aids in enhancing the reconstruction of textures and details.

\subsection{Overall Fine-tuning Procedure}
\label{Overview fine-tuning process}
Fig.~\ref{fig:MoE-DiffIR}(a) illustrates the entire fine-tuning process of our MoE-DiffIR. Similar to StableSR~\cite{wang2023exploiting-stablesr} and AutoDIR~\cite{jiang2023autodir}, we fine-tune the framework in two stages. In the first stage, only the MoE-Prompt Module is trained, while the VAE Codecs and UNet remain fixed. The MoE-Prompt Module modulates the LQ image features onto the multi-scale outputs of Stable Diffusion via the SPADE layer~\cite{wang2018recovering-spade}. To achieve this, we employ three downsample layers in the MoE-Prompt Module, and use ViT blocks~\cite{dosovitskiy2020image-VIT} and convolution layers to extract LQ features at each scale.

In the second stage, all modules are fixed except for the VAE decoder. This fine-tuning process is crucial for ensuring the fidelity of the recovered images, which is also underscored in existing literature~\cite{zhu2023designing-decoder-compensator,wang2023exploiting-stablesr}. The high compression rate may lead to information loss during the reconstruction stage via the VAE decoder. This occurs because the pre-trained VAE decoder does not align well with varying scenarios, causing the output latent variable $z_{0}$ from Stable Diffusion to misalign with the our CIR tasks. Consequently, it is essential to augment the Decoder with some low-quality information, as clearly illustrated in Fig.~\ref{fig:MoE-DiffIR}(d). The loss function for second stage fine-tuning is:
\begin{equation}
    L_{Decoder} = \mathcal{L}_{lpips}[z_{lq},z{_0},hr]
\end{equation}
In this phase, we employ the LPIPS perceptual loss function, using high-quality images as the reference. Here $z_{0}$ denotes the output of unet denoising network and $z_{lq}$ is latent variable of low quality image.

\subsection{CIR dataset benchmark}
\label{CIR dataset benchmark}
We introduce the first universal dataset benchmark for compressed image restoration. This benchmark includes four traditional compression methods: (i) JPEG~\cite{wallace1991jpeg}, (ii) VVC~\cite{bross2021overview-VVC}, (iii) HEVC~\cite{sullivan2012overview-HEVC}, (iv) WebP~\cite{ginesu2012objective-WEBP} and three learning-based compression methods: (i) $C_{PSNR}$, (ii) $C_{SSIM}$, (iii) HIFIC~\cite{mentzer2020high-HIFIC}. Both $C_{PSNR}$ and $C_{SSIM}$ are adopted from the work~\cite{cheng2020learned}, optimized by MSE and MS-SSIM loss, respectively. Each codec has three distinct bitrate levels. For JPEG and WebP, we set quality factor (QF) values from [5,10,15]. For VVC and HEVC, we adopt MPEG standard QP values from [37, 42, 47]. For HIFIC, we use three released weights~\footnote{\url{https://github.com/Justin-Tan/high-fidelity-generative-compression}} represented for three different bitrates: ``low'', ``med'', and ``high''. We also define cross-degree distortions for unseen test tasks, such as setting QF of JPEG from values [5,25]. Additionally, we create cross-type distortions using AVC codec methods for static images from values [37, 42, 47]. We adopt DF2K~\cite{agustsson2017ntire-DIV2K,lim2017enhanced} as our compressed training dataset, containing 3450 images, resulting in 72450 images across 21 compression tasks.

\section{Experiments}
\label{sec:Experiments}
\subsection{Experiment Setup}

\noindent\textbf{Implementation Details.}
We fine-tune Stable Diffusion 2.1-base\footnote{\url{https://huggingface.co/stabilityai/stable-diffusion-2-1-base}} over two training stages. In the first stage, followed in Sec.~\ref{Overview fine-tuning process}, we fix the decoder of VAE and only train MoE-Prompt module. We use an Adam optimizer($\beta_{1}=0.9$, $\beta_{2}=0.999$) with fixed learning rate of $5e^{-5}$. The total iterations are 0.4M steps, constrained by loss function $\mathcal{L}_{SD}$  as described in the Sec.~\ref{preliminary}. In the second stage, we train only the decoder compensator with other modules fixed.  We generate 70,000 latent images using the weights from the first stage and train the decoder with the corresponding LQ images and ground truth images. The learning rate is set to $1e^{-4}$ and total iterations are 0.1M steps. In the whole training process, we resize the input images into 256x256 and employ random flipping and rotation for data augmentation. The batch size is set to 32, and the training is conducted on four NVIDIA RTX 3090 GPUs.

\noindent\textbf{Compared Methods.}
To validate the effectiveness of MoE-DiffIR, we compare it with several state-of-the-art (SOTA) methods. These methods include two for all-in-one IR: PromptIR~\cite{potlapalli2023promptir} and Airnet~\cite{li2022all-AirNet}, one method for compression artifact removal: HAT~\cite{chen2023hat}, one GAN-based method: RealESRGAN~\cite{wang2021realesrgan}, and four diffusion-based methods: StableSR~\cite{wang2023exploiting-stablesr}, DiffBIR~\cite{lin2023diffbir}, SUPIR~\cite{yu2024scaling-SUPIR} and PASD~\cite{yang2023pixel-PASD}. Here, we present only a subset of the quantitative results. A more comprehensive set of quantitative results will be detailed in the Sec.~\ref{Appendix:more quantitative studies} of the \textbf{Appendix}. For training settings, we adhere to the configurations provided in the official code repositories of these methods. We set batch size to 32 for all methods.
\subsection{Comparisons with State-of-the-arts}
We validate MoE-DiffIR on five commonly used compressed test sets: LIVE1~\cite{sheikh2005live1}, Classic5~\cite{zeyde2012single-classic5}, BSDS500~\cite{arbelaez2010contour-BSDS}, DIV2K Testset~\cite{agustsson2017ntire-DIV2K}, and ICB~\cite{ehrlich2020quantization_ICB}. We employ PSNR, SSIM as distortions metrics, and LPIPS~\cite{zhang2018unreasonable-LPIPS}, FID~\cite{heusel2017gans-FID} as perceptual metrics. In Sec.~\ref{Appendix:more quantitative studies} of the \textbf{Appendix}, we show more results using some non-reference metrics like ClipIQA~\cite{wang2023exploring-CLIPIQA}, ManIQA~\cite{yang2022maniqa} to further validate the perceptual quality.

\clearpage

\begin{table*}[htp]
\centering
\caption{Quantitative comparison for compressed image restoration on seven codecs (average on three distortions). Results are tested on with different compression qualities in terms of  PSNR$\uparrow$, SSIM$\uparrow$, LPIPS$\downarrow$, FID$\downarrow$. \textcolor{red}{Red} and \textcolor{blue}{blue} colors represent the best and second best performance, respectively. All comparison methods are reproduced on our constructed CIR datasets. (Suggested to zoom in for better visualization.)}
\setlength{\tabcolsep}{2pt}
\resizebox{\textwidth}{!}{
\renewcommand{\arraystretch}{0.95}
\begin{tabular}{c|c|cccc|cccc|cccc|cccc|cccc}
\Xhline{2pt}
                          &                           & \multicolumn{4}{c|}{LIVE1~\cite{sheikh2005live1}}                                                                                                  & \multicolumn{4}{c|}{Classic5~\cite{zeyde2012single-classic5}}                                                                                               & \multicolumn{4}{c|}{BSDS500~\cite{arbelaez2010contour-BSDS}}                                                                                                & \multicolumn{4}{c|}{DIV2K~\cite{agustsson2017ntire-DIV2K}}                                                                                                 & \multicolumn{4}{c}{ICB~\cite{ehrlich2020quantization_ICB}}                                                                                                     \\ \cline{3-22} 
\multirow{-2}{*}{Codecs}  & \multirow{-2}{*}{Methods} & PSNR                         & SSIM                         & LPIPS                         & FID                           & PSNR                         & SSIM                         & LPIPS                         & FID                           & PSNR                         & SSIM                         & LPIPS                         & FID                           & PSNR                         & SSIM                         & LPIPS                         & FID                          & PSNR                         & SSIM                         & LPIPS                         & FID                           \\
\end{tabular}
}
\label{table:traditional}
\end{table*}

\noindent\textbf{Quantitative Analysis.} Table~\ref{table:traditional} shows comprehensive performance of our MoE-DiffIR compared with SOTA methods across 7 compression codecs. Here, for each codec, we average the metrics of its three distortion levels. Since the primary objective of this work is to enhance the perceptual quality of images at low bitrates, our comparisons primarily focus on the perceptual quality against generative models. From the table, we can see that our method almost surpasses all other methods in terms of perceptual metrics like LPIPS and FID. Moreover, our method is also competitive in distortion metrics such as PSNR compared to transformer-based methods, thanks to the fine-tuning stage of the VAE decoder. Specifically, on the LPIPS metric, we achieved a 10.9\% reduction compared to SUPIR, a decrease of 5.4 on the FID metric, and also an average increase of 0.41dB over StableSR on the PSNR metric.

\begin{figure}[H]
	\centering
	\includegraphics[width=1\linewidth]{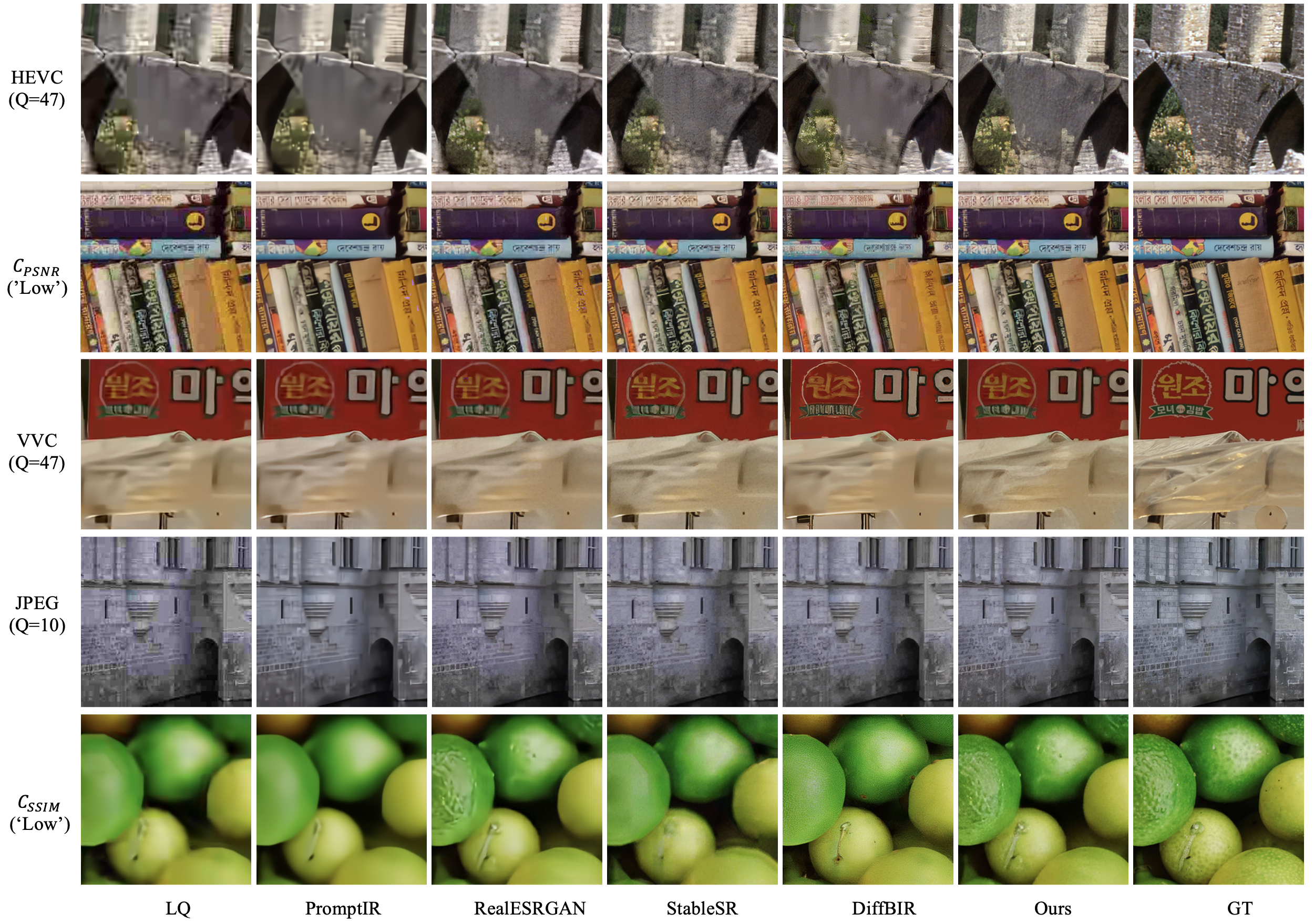}
	\caption{Visual comparisons between our methods and other state of the arts methods. This figure demonstrate 5 different compression tasks: JPEG (QF=10), VVC (QP=47), HEVC (QP=47), $C_{SSIM}$(``Low'' bitrates), $C_{PSNR}$(``Low'' bitrates). More visual results can be found in Sec.~\ref{Appendix:more visual Results} of the \textbf{Appendix}.}
	\label{fig:MoE-DiffIR-results}
\end{figure}

\noindent \textbf{Qualitative Analysis.} Additionally, we also present some perceptual visual results in Fig.~\ref{fig:MoE-DiffIR-results}, covering different quality factors from various codecs (More visual comparsions with SUPIR and PASD are shown in Sec.~\ref{Appendix:more visual Results} of the \textbf{Appendix}). It is observable that, in scenarios with lower compression rates, the Transformer-based all-in-one model PromptIR tends to restore images too smoothly, whereas DiffBIR is prone to generating some erroneous texture details, as shown in the textual information in the third row of Fig.~\ref{fig:MoE-DiffIR-results}. Thanks to the compensation in the second stage of the VAE decoder, MoE-DiffIR is capable of generating more accurate textures in terms of fidelity. Moreover, our MoE-Prompt enables MoE-DiffIR to effectively handle different compression distortions, demonstrating excellent perceptual restoration capabilities, including color correction and texture detail generation.

\subsection{Ablation Studies}
\label{subsec:ablation study}
\subsubsection{The effects of MoE-Prompt.}
We conduct experiments with different prompt designs mentioned in Sec.~~\ref{sec:Methodology}. The results are presented in Table~\ref{table:ablation1}. We compare four prompt designs: No Prompt, Single Prompt, Multiple Prompts, and our MoE-Prompt. In addition to the 21 tasks in our CIR dataset, we also test the performance of these models on some unseen tasks to assess their generalizability. Specifically, we employ two types of unseen tasks to validate the generalization performance. The first type is ``Cross Degrees'', which involves selecting one of the seven codecs, but with unseen quality factors. In this experiment, we choose VVC with QP from values [32,52] as the distortion types. The second type is ``Cross Type'', where we select codec AVC~\cite{wiegand2003overview-H.264} with QP from values [47, 42, 37]. Specifically, without prompts, the model has a reduced ability to distinguish between various distortions, leading to a notably lower average performance across tasks than prompt-based models, particularly on unseen tasks. Furthermore, using a single prompt results in lower average performance across the 21 tasks compared to using multiple prompts or MoE-Prompt, indicating that a single prompt lacks the capability for multi-task modeling. In contrast, our MoE-based method exceeds multiple prompt design by an average of 0.05dB on 21 tasks and improves perceptual quality, reducing LPIPS by 5\% and FID by about 4. This proves that MoE-Prompt can more effectively utilize and share prompts across various distortion tasks, uncovering task-customized diffusion priors than other prompt interaction methods.

\begin{table*}[htbp]
\centering
\caption{Impacts of different prompt designs. Results are reported on LIVE1, DIV2K and ICB. For seen tasks, the value is the average result of 21 compression tasks on our CIR dataset. For unseen tasks, the value is the average result of ``Cross Degrees'' (VVC: QP form values [32,52]) and ``Cross Types'' (AVC: QP from values [47,42,37]).  Best performances are in \textbf{bold}.}
\resizebox{\textwidth}{!}{
\setlength{\tabcolsep}{1.5pt}
\begin{tabular}{c|cccccc|clclclcl}
\Xhline{2pt}
                          & \multicolumn{6}{c|}{Seen tasks}                                                                                                                                                      & \multicolumn{8}{c}{Unseen Tasks}                                                                                                                                                                                                                                          \\ \cline{2-15} 
                          & \multicolumn{2}{c|}{LIVE1}                                        & \multicolumn{2}{c|}{BSDS500}                                      & \multicolumn{2}{c|}{DIV2K}                   & \multicolumn{4}{c|}{LIVE1 (Cross Degrees)}                                                                                           & \multicolumn{4}{c}{ICB (Cross Types)}                                                                                                \\ \cline{2-15} 
\multirow{-3}{*}{Methods} & PSNR/SSIM            & \multicolumn{1}{c|}{LPIPS/FID}             & PSNR/SSIM            & \multicolumn{1}{c|}{LPIPS/FID}             & PSNR/SSIM            & LPIPS/FID             & \multicolumn{2}{c}{PSNR/SSIM}                                   & \multicolumn{2}{c|}{LPIPS/FID}                                    & \multicolumn{2}{c}{PSNR/SSIM}                                   & \multicolumn{2}{c}{LPIPS/FID}                                     \\ \hline
No Prompt                 & 28.73/0.806          & \multicolumn{1}{c|}{0.1343/85.87}          & 28.56/0.770          & \multicolumn{1}{c|}{0.1591/96.45}          & 27.86/0.813          & 0.1295/79.5           & \multicolumn{2}{c}{31.79/0.893}                                 & \multicolumn{2}{c|}{0.064/37.33}                                  & \multicolumn{2}{c}{28.61/0.787}                                 & \multicolumn{2}{c}{0.1933/210.84}                                 \\
Single Prompt             & 28.86/0.806          & \multicolumn{1}{c|}{0.1272/79.45}          & 28.78/0.791          & \multicolumn{1}{c|}{0.1530/89.62}          & 28.02/0.816          & 0.1143/71.26          & \multicolumn{2}{c}{33.25/0.910}                                 & \multicolumn{2}{c|}{0.0457/28.60}                                 & \multicolumn{2}{c}{28.88/0.793}                                 & \multicolumn{2}{c}{0.179/187.29}                                  \\
Multiple Prompt           & 28.98/0.810          & \multicolumn{1}{c|}{0.1212/77.09}          & 28.93/0.794          & \multicolumn{1}{c|}{0.1482/89.34}          & 28.22/0.817          & 0.1124/71.65          & \multicolumn{2}{c}{33.32/0.913}                                 & \multicolumn{2}{c|}{0.0432/28.23}                                 & \multicolumn{2}{c}{28.89/0.792}                                 & \multicolumn{2}{c}{0.1756/187.89}                                 \\
 MoE-Prompt (Ours)          & \textbf{29.02/0.811} & \multicolumn{1}{c|}{\textbf{0.1179/75.86}} & \textbf{28.97/0.794} & \multicolumn{1}{c|}{\textbf{0.1430/88.14}} & \textbf{28.29/0.821} & \textbf{0.1071/68.91} & \multicolumn{2}{c}{{\color[HTML]{000000} \textbf{33.45/0.916}}} & \multicolumn{2}{c|}{{\color[HTML]{000000} \textbf{0.0411/25.65}}} & \multicolumn{2}{c}{{\color[HTML]{000000} \textbf{29.02/0.800}}} & \multicolumn{2}{c}{{\color[HTML]{000000} \textbf{0.1690/176.87}}}\\ \Xhline{2pt}
\end{tabular}
}
\label{table:ablation1}
\end{table*}
\setlength{\intextsep}{7pt}
\begin{table*}[htbp]
\centering
\caption{Impacts of Visual2Text(V2T) adapter and Degradation Prior (DP). Results are reported on LIVE1, BSDS500, ICB. Here the value is the average result of 21 compression tasks. Best performances are in \textbf{bold}.}
\resizebox{\textwidth}{!}{
\renewcommand{\arraystretch}{0.6}
\setlength{\tabcolsep}{9pt}
\begin{tabular}{c|cccccc}
\Xhline{2pt}
\multirow{3}{*}{Methods}    & \multicolumn{6}{c}{Datasets}                                                                                                    \\ \cline{2-7} 
                            & \multicolumn{2}{c|}{LIVE1}                      & \multicolumn{2}{c|}{BSDS500}                    & \multicolumn{2}{c}{ICB}     \\ \cline{2-7} 
                            & PSNR/SSIM   & \multicolumn{1}{c|}{LPIPS/FID}    & PSNR/SSIM   & \multicolumn{1}{c|}{LPIPS/FID}    & PSNR/SSIM   & LPIPS/FID     \\ \hline
MoE-Prompt                 & 29.02/0.810 & \multicolumn{1}{c|}{0.1179/75.86} & 28.97/0.794 & \multicolumn{1}{c|}{0.1430/88.14} & 29.83/0.839 & 0.1277/122.59 \\ \hline
MoE-Prompt+V2T Adapter    & 29.03/0.812 & \multicolumn{1}{c|}{0.1145/74.13} & 28.94/0.796 & \multicolumn{1}{c|}{0.1367/\textbf{86.77}} & 29.83/0.840 & 0.1239/119.78 \\ \hline
MoE-Prompt+DP              & 29.07/0.814 & \multicolumn{1}{c|}{0.1154/76.60} & \textbf{29.06}/0.795 & \multicolumn{1}{c|}{0.1405/88.00} & 29.87/0.841 & 0.1269/122.32 \\ \hline
MoE-Prompt+V2T Adapter+DP & \textbf{29.10/0.814} & \multicolumn{1}{c|}{\textbf{0.1136/73.60}} & 29.02/\textbf{0.797} & \multicolumn{1}{c|}{\textbf{0.1356}/86.81} & \textbf{29.88/0.841} & \textbf{0.1235/119.29} \\ \Xhline{2pt}
\end{tabular}
}
\label{table:ablation2}
\end{table*}

\subsubsection{The effects of Visual2Text adapter and Degradation Prior.}
\begin{figure}[htp]
	\centering
	\includegraphics[width=1\linewidth]{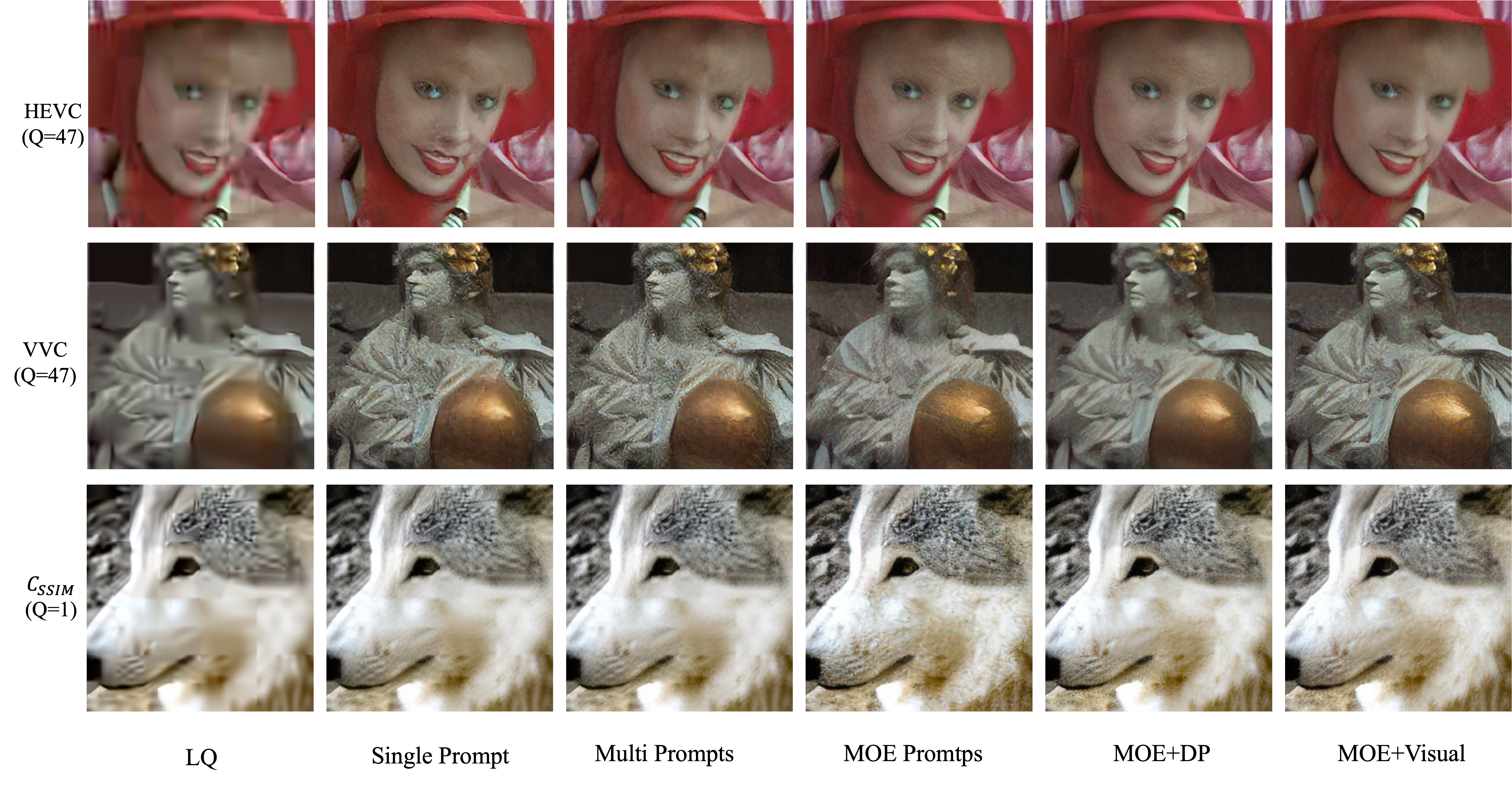}
	\caption{Visual ablation results: different prompt interaction designs, use of V2T adapter and use of degradation prior (DP).}
	\label{fig:Ablation1-2}
\end{figure}

In Sec.~\ref{Visual2Text}, we describe using a cross-modal adapter to convert visual information into text embeddings. Additionally, in Sec.~\ref{MoE Prompts Methodology}, we employ the pre-trained DACLIP~\cite{luo2023controlling-DACLIP} to provide degradation priors (DP), enhancing the router's adaptive selection of optimal prompts. Ablation studies validate these methodologies by integrating the Visual2Text adapter or DP into the MoE-Prompt backbone. Table~\ref{table:ablation2} shows that adding V2T adapter could reduce LPIPS by 3-5\% and improves FID by 1-3 points on average, indicating better perceptual quality. The use of degradation prior (DP) mainly contributes to distortion metrics, with an average PSNR increase of 0.07dB across 21 tasks, indicating that adding visual information could enhance perceptual quality while adding DP may improve fidelity. Visual comparisons in Fig.~\ref{fig:Ablation1-2} also show an interesting phenomenon: at extremely low bitrates, Stable Diffusion may convert severe distortions into noise spots, which could be smoothed with the use of the V2T adapter or by adding degradation prior (DP), thereby enhancing model performance.

\begin{figure}[htbp]
\centering
\begin{subfigure}[b]{0.45\textwidth}
    \includegraphics[width=\textwidth]{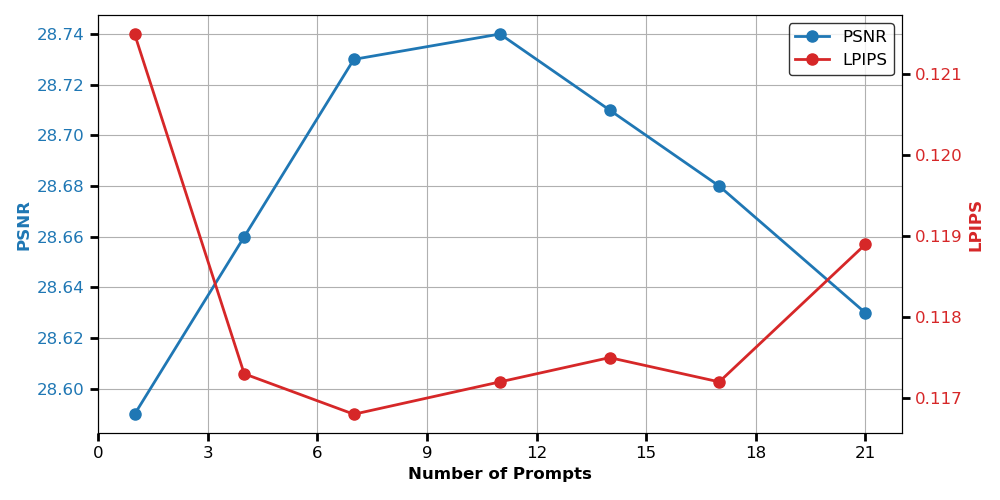}
    \caption{Effects of Number of Prompts}
    \label{fig:image1}
\end{subfigure}
\hspace{1mm}
\begin{subfigure}[b]{0.45\textwidth}
    \includegraphics[width=\textwidth]{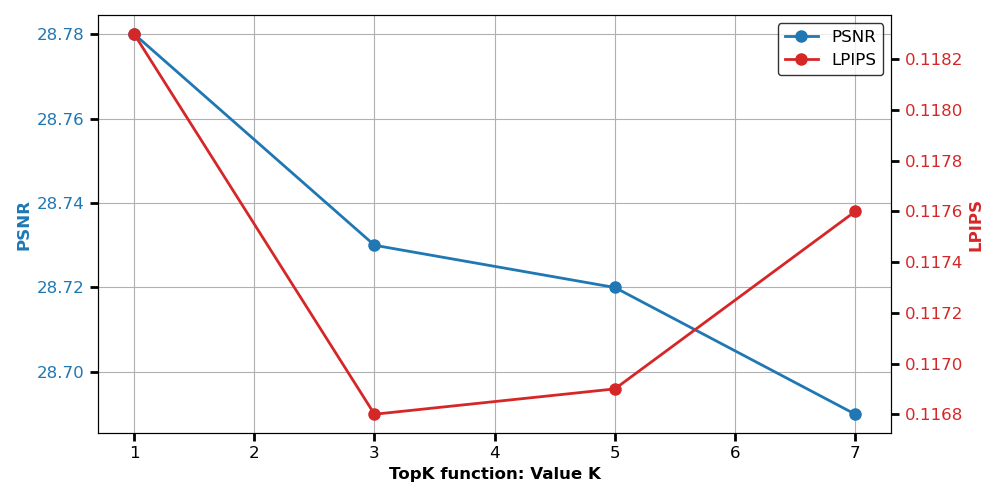}
    \caption{Effects of K value}
    \label{fig:image2}
\end{subfigure}
\caption{The effect of the number of prompts and the value $K$ of the Top-K function.}
\label{fig:Ablation3}
\end{figure}

\subsubsection{The effects of number of total prompts and selected prompts.}
In our proposed MoE-Prompt Module, the router uses a Top-K function to select $K$ prompts from $N$ predefined prompts. We conduct ablation experiments to evaluate the effects of varying $N$ and $K$. Since our CIR dataset consits of total 21 compression tasks, we set $N$ to a series of values (1, 4, 7, 11, 14, 17, 21) and use the LIVE1 dataset for testing. As shown in Fig.~\ref{fig:Ablation3}(a), changing $N$ significantly impacts both the distortion metric PSNR and the perceptual metric LPIPS. When $N$ is small, both PSNR and LPIPS are poor because a single prompt ($N$=1) cannot model different distortion tasks effectively. As $N$ increases, performance improves but then declines past a certain point due to the difficulty in learning task-relevant distortion features from too many prompts, leading to underutilization. The data suggests that performance is optimal at $N$=7, similar to $N$=11, indicating that around $N$=7 is sufficient for parameter economy. We then fix $N$ at 7 and vary $K$ with values (1, 3, 5, 7). Fig.~\ref{fig:Ablation3}(b) shows that while $K$=1 yields higher PSNR, it results in poor perceptual quality. We can conclude that multiple prompts could cooperate together for perceiving different tasks with better perceptual quality. The best perceptual performance is seen when $K$ is between 3 and 5. Thus, we select $N$=7 and $K$=3 for the final settings.

\section{Conclusion}
In this work, we propose the MoE-Prompt to excavate task-customized diffusion priors for universal compressed image restoration, dubbed MoE-DiffIR. Our method maximizes the utilization of different prompts, enabling them to collaboratively perceive different distortions. By utilizing a Visual2Text adapter, we integrate visual information into the text inputs of the Stable Diffusion model, thereby improving the perceptual restoration capabilities of the model at low bitrates. We also construct a comprehensive dataset benchmark for CIR tasks. Our extensive experiments have demonstrated that MoE-DiffIR not only improves perceptual performance at low bitrates but also facilitates rapid transferability across various compression tasks. In the future, we intend to design novel approaches within our CIR benchmark to further improve the performance of the model.

\section*{Limitation}
In this work, we propose a novel universal compressed image restoration (CIR) framework using the MoE-Prompt Modules. Although our model outperforms other methods in terms of perceptual quality, there remains a notable gap between the restored images and the ground truth at extremely low bitrates, as shown in Fig.~\ref{fig:MoE-DiffIR-results}. In future work, we aim to focus on this area and strive for improvements.
\section*{Acknowledgements}
This work was supported in part by NSFC under Grant 623B2098, 62371434 and 62021001.
\clearpage 
\bibliographystyle{splncs04}
\bibliography{egbib}

\begin{thebibliography}{10}
\providecommand{\url}[1]{\texttt{#1}}
\providecommand{\urlprefix}{URL }
\providecommand{\doi}[1]{https://doi.org/#1}

\bibitem{aggarwal2001surprising-cos1}
Aggarwal, C.C., Hinneburg, A., Keim, D.A.: On the surprising behavior of distance metrics in high dimensional space. In: Database Theory—ICDT 2001: 8th International Conference London, UK, January 4--6, 2001 Proceedings 8. pp. 420--434. Springer (2001)

\bibitem{agustsson2023multi_compression_new4}
Agustsson, E., Minnen, D., Toderici, G., Mentzer, F.: Multi-realism image compression with a conditional generator. In: Proceedings of the IEEE/CVF Conference on Computer Vision and Pattern Recognition. pp. 22324--22333 (2023)

\bibitem{agustsson2017ntire-DIV2K}
Agustsson, E., Timofte, R.: Ntire 2017 challenge on single image super-resolution: Dataset and study. In: Proceedings of the IEEE conference on computer vision and pattern recognition workshops. pp. 126--135 (2017)

\bibitem{ai2023multimodal-mperceiver}
Ai, Y., Huang, H., Zhou, X., Wang, J., He, R.: Multimodal prompt perceiver: Empower adaptiveness, generalizability and fidelity for all-in-one image restoration. arXiv preprint arXiv:2312.02918  (2023)

\bibitem{arbelaez2010contour-BSDS}
Arbelaez, P., Maire, M., Fowlkes, C., Malik, J.: Contour detection and hierarchical image segmentation. IEEE transactions on pattern analysis and machine intelligence  \textbf{33}(5),  898--916 (2010)

\bibitem{bross2021overview-VVC}
Bross, B., Wang, Y.K., Ye, Y., Liu, S., Chen, J., Sullivan, G.J., Ohm, J.R.: Overview of the versatile video coding (vvc) standard and its applications. IEEE Transactions on Circuits and Systems for Video Technology  \textbf{31}(10),  3736--3764 (2021)

\bibitem{pyiqa}
Chen, C., Mo, J.: {IQA-PyTorch}: Pytorch toolbox for image quality assessment. [Online]. Available: \url{https://github.com/chaofengc/IQA-PyTorch} (2022)

\bibitem{chen2023hat}
Chen, X., Wang, X., Zhang, W., Kong, X., Qiao, Y., Zhou, J., Dong, C.: Hat: Hybrid attention transformer for image restoration. arXiv preprint arXiv:2309.05239  (2023)

\bibitem{cheng2020learned}
Cheng, Z., Sun, H., Takeuchi, M., Katto, J.: Learned image compression with discretized gaussian mixture likelihoods and attention modules. In: Proceedings of the IEEE/CVF conference on computer vision and pattern recognition. pp. 7939--7948 (2020)

\bibitem{christopoulos2000jpeg2000}
Christopoulos, C., Skodras, A., Ebrahimi, T.: The jpeg2000 still image coding system: an overview. IEEE transactions on consumer electronics  \textbf{46}(4),  1103--1127 (2000)

\bibitem{chung2022diffusion-dps}
Chung, H., Kim, J., Mccann, M.T., Klasky, M.L., Ye, J.C.: Diffusion posterior sampling for general noisy inverse problems. arXiv preprint arXiv:2209.14687  (2022)

\bibitem{chung2023prompt-P2L}
Chung, H., Ye, J.C., Milanfar, P., Delbracio, M.: Prompt-tuning latent diffusion models for inverse problems. arXiv preprint arXiv:2310.01110  (2023)

\bibitem{dong2015compression-Learning-based1}
Dong, C., Deng, Y., Loy, C.C., Tang, X.: Compression artifacts reduction by a deep convolutional network. In: Proceedings of the IEEE international conference on computer vision. pp. 576--584 (2015)

\bibitem{dosovitskiy2020image-VIT}
Dosovitskiy, A., Beyer, L., Kolesnikov, A., Weissenborn, D., Zhai, X., Unterthiner, T., Dehghani, M., Minderer, M., Heigold, G., Gelly, S., et~al.: An image is worth 16x16 words: Transformers for image recognition at scale. arXiv preprint arXiv:2010.11929  (2020)

\bibitem{dubey2016cosine-cos2}
Dubey, V.K., Saxena, A.K.: Cosine similarity based filter technique for feature selection. In: 2016 International Conference on Control, Computing, Communication and Materials (ICCCCM). pp.~1--6. IEEE (2016)

\bibitem{ehrlich2020quantization-QGAC}
Ehrlich, M., Davis, L., Lim, S.N., Shrivastava, A.: Quantization guided jpeg artifact correction. In: Computer Vision--ECCV 2020: 16th European Conference, Glasgow, UK, August 23--28, 2020, Proceedings, Part VIII 16. pp. 293--309. Springer (2020)

\bibitem{ehrlich2020quantization_ICB}
Ehrlich, M., Davis, L., Lim, S.N., Shrivastava, A.: Quantization guided jpeg artifact correction. In: Computer Vision--ECCV 2020: 16th European Conference, Glasgow, UK, August 23--28, 2020, Proceedings, Part VIII 16. pp. 293--309. Springer (2020)

\bibitem{foi2007pointwise-model-based1}
Foi, A., Katkovnik, V., Egiazarian, K.: Pointwise shape-adaptive dct for high-quality denoising and deblocking of grayscale and color images. IEEE transactions on image processing  \textbf{16}(5),  1395--1411 (2007)

\bibitem{gao2024clip-adapter}
Gao, P., Geng, S., Zhang, R., Ma, T., Fang, R., Zhang, Y., Li, H., Qiao, Y.: Clip-adapter: Better vision-language models with feature adapters. International Journal of Computer Vision  \textbf{132}(2),  581--595 (2024)

\bibitem{ginesu2012objective-WEBP}
Ginesu, G., Pintus, M., Giusto, D.D.: Objective assessment of the webp image coding algorithm. Signal processing: image communication  \textbf{27}(8),  867--874 (2012)

\bibitem{gou2023exploiting-all-in-one}
Gou, Y., Zhao, H., Li, B., Xiao, X., Peng, X.: Exploiting diffusion priors for all-in-one image restoration. arXiv preprint arXiv:2312.02197  (2023)

\bibitem{he2022elic-compression_new2}
He, D., Yang, Z., Peng, W., Ma, R., Qin, H., Wang, Y.: Elic: Efficient learned image compression with unevenly grouped space-channel contextual adaptive coding. In: Proceedings of the IEEE/CVF Conference on Computer Vision and Pattern Recognition. pp. 5718--5727 (2022)

\bibitem{heusel2017gans-FID}
Heusel, M., Ramsauer, H., Unterthiner, T., Nessler, B., Hochreiter, S.: Gans trained by a two time-scale update rule converge to a local nash equilibrium. Advances in neural information processing systems  \textbf{30} (2017)

\bibitem{jia2022visual-prompt-latest3}
Jia, M., Tang, L., Chen, B.C., Cardie, C., Belongie, S., Hariharan, B., Lim, S.N.: Visual prompt tuning. In: European Conference on Computer Vision. pp. 709--727. Springer (2022)

\bibitem{jiang2021towards-FBCNN}
Jiang, J., Zhang, K., Timofte, R.: Towards flexible blind jpeg artifacts removal. In: Proceedings of the IEEE/CVF International Conference on Computer Vision. pp. 4997--5006 (2021)

\bibitem{jiang2023autodir}
Jiang, Y., Zhang, Z., Xue, T., Gu, J.: Autodir: Automatic all-in-one image restoration with latent diffusion. arXiv preprint arXiv:2310.10123  (2023)

\bibitem{jin2021dc}
Jin, Y., Sharma, A., Tan, R.T.: Dc-shadownet: Single-image hard and soft shadow removal using unsupervised domain-classifier guided network. In: Proceedings of the IEEE/CVF International Conference on Computer Vision. pp. 5027--5036 (2021)

\bibitem{jin2024des3}
Jin, Y., Ye, W., Yang, W., Yuan, Y., Tan, R.T.: Des3: Adaptive attention-driven self and soft shadow removal using vit similarity. In: Proceedings of the AAAI Conference on Artificial Intelligence. vol.~38, pp. 2634--2642 (2024)

\bibitem{kawar2022denoising-ddrm}
Kawar, B., Elad, M., Ermon, S., Song, J.: Denoising diffusion restoration models. Advances in Neural Information Processing Systems  \textbf{35},  23593--23606 (2022)

\bibitem{krizhevsky2012imagenet-CNN}
Krizhevsky, A., Sutskever, I., Hinton, G.E.: Imagenet classification with deep convolutional neural networks. Advances in neural information processing systems  \textbf{25} (2012)

\bibitem{li2024promptcir}
Li, B., Li, X., Lu, Y., Feng, R., Guo, M., Zhao, S., Zhang, L., Chen, Z.: Promptcir: Blind compressed image restoration with prompt learning. Proceedings of the IEEE/CVF Conference on Computer Vision and Pattern Recognition Workshops  (2024)

\bibitem{li2024sed}
Li, B., Li, X., Zhu, H., Jin, Y., Feng, R., Zhang, Z., Chen, Z.: Sed: Semantic-aware discriminator for image super-resolution. In: Proceedings of the IEEE/CVF Conference on Computer Vision and Pattern Recognition. pp. 25784--25795 (2024)

\bibitem{li2022all-AirNet}
Li, B., Liu, X., Hu, P., Wu, Z., Lv, J., Peng, X.: All-in-one image restoration for unknown corruption. In: Proceedings of the IEEE/CVF Conference on Computer Vision and Pattern Recognition. pp. 17452--17462 (2022)

\bibitem{li2022srdiff}
Li, H., Yang, Y., Chang, M., Chen, S., Feng, H., Xu, Z., Li, Q., Chen, Y.: Srdiff: Single image super-resolution with diffusion probabilistic models. Neurocomputing  \textbf{479},  47--59 (2022)

\bibitem{li2020learning-disentangle}
Li, X., Jin, X., Lin, J., Liu, S., Wu, Y., Yu, T., Zhou, W., Chen, Z.: Learning disentangled feature representation for hybrid-distorted image restoration. In: Computer Vision--ECCV 2020: 16th European Conference, Glasgow, UK, August 23--28, 2020, Proceedings, Part XXIX 16. pp. 313--329. Springer (2020)

\bibitem{li2023diffusion-survey}
Li, X., Ren, Y., Jin, X., Lan, C., Wang, X., Zeng, W., Wang, X., Chen, Z.: Diffusion models for image restoration and enhancement--a comprehensive survey. arXiv preprint arXiv:2308.09388  (2023)

\bibitem{li2021task-code-task}
Li, X., Shi, J., Chen, Z.: Task-driven semantic coding via reinforcement learning. IEEE Transactions on Image Processing  \textbf{30},  6307--6320 (2021)

\bibitem{li2020multi-vvc}
Li, X., Sun, S., Zhang, Z., Chen, Z.: Multi-scale grouped dense network for vvc intra coding. In: Proceedings of the IEEE/CVF Conference on Computer Vision and Pattern Recognition Workshops. pp. 158--159 (2020)

\bibitem{li2023prompt-PIP}
Li, Z., Lei, Y., Ma, C., Zhang, J., Shan, H.: Prompt-in-prompt learning for universal image restoration. arXiv preprint arXiv:2312.05038  (2023)

\bibitem{liang2021swinir}
Liang, J., Cao, J., Sun, G., Zhang, K., Van~Gool, L., Timofte, R.: Swinir: Image restoration using swin transformer. In: Proceedings of the IEEE/CVF international conference on computer vision. pp. 1833--1844 (2021)

\bibitem{lim2017enhanced}
Lim, B., Son, S., Kim, H., Nah, S., Mu~Lee, K.: Enhanced deep residual networks for single image super-resolution. In: Proceedings of the IEEE conference on computer vision and pattern recognition workshops. pp. 136--144 (2017)

\bibitem{lin2023diffbir}
Lin, X., He, J., Chen, Z., Lyu, Z., Fei, B., Dai, B., Ouyang, W., Qiao, Y., Dong, C.: Diffbir: Towards blind image restoration with generative diffusion prior. arXiv preprint arXiv:2308.15070  (2023)

\bibitem{luo2023image-MoE3}
Luo, F., Xiang, J., Zhang, J., Han, X., Yang, W.: Image super-resolution via latent diffusion: A sampling-space mixture of experts and frequency-augmented decoder approach. arXiv preprint arXiv:2310.12004  (2023)

\bibitem{luo2023controlling-DACLIP}
Luo, Z., Gustafsson, F.K., Zhao, Z., Sj{\"o}lund, J., Sch{\"o}n, T.B.: Controlling vision-language models for universal image restoration. arXiv preprint arXiv:2310.01018  (2023)

\bibitem{luo2023image-IRSDE}
Luo, Z., Gustafsson, F.K., Zhao, Z., Sj{\"o}lund, J., Sch{\"o}n, T.B.: Image restoration with mean-reverting stochastic differential equations. arXiv preprint arXiv:2301.11699  (2023)

\bibitem{luo2023refusion}
Luo, Z., Gustafsson, F.K., Zhao, Z., Sj{\"o}lund, J., Sch{\"o}n, T.B.: Refusion: Enabling large-size realistic image restoration with latent-space diffusion models. In: Proceedings of the IEEE/CVF Conference on Computer Vision and Pattern Recognition. pp. 1680--1691 (2023)

\bibitem{ma2023prores}
Ma, J., Cheng, T., Wang, G., Zhang, Q., Wang, X., Zhang, L.: Prores: Exploring degradation-aware visual prompt for universal image restoration. arXiv preprint arXiv:2306.13653  (2023)

\bibitem{masoudnia2014mixture-MoE1}
Masoudnia, S., Ebrahimpour, R.: Mixture of experts: a literature survey. Artificial Intelligence Review  \textbf{42},  275--293 (2014)

\bibitem{mentzer2020high-HIFIC}
Mentzer, F., Toderici, G.D., Tschannen, M., Agustsson, E.: High-fidelity generative image compression. Advances in Neural Information Processing Systems  \textbf{33},  11913--11924 (2020)

\bibitem{moser2024diffusion-survey-2}
Moser, B.B., Shanbhag, A.S., Raue, F., Frolov, S., Palacio, S., Dengel, A.: Diffusion models, image super-resolution and everything: A survey. arXiv preprint arXiv:2401.00736  (2024)

\bibitem{mou2023t2i}
Mou, C., Wang, X., Xie, L., Wu, Y., Zhang, J., Qi, Z., Shan, Y., Qie, X.: T2i-adapter: Learning adapters to dig out more controllable ability for text-to-image diffusion models. arXiv preprint arXiv:2302.08453  (2023)

\bibitem{nie2023pro-prompt-latest1}
Nie, X., Ni, B., Chang, J., Meng, G., Huo, C., Xiang, S., Tian, Q.: Pro-tuning: Unified prompt tuning for vision tasks. IEEE Transactions on Circuits and Systems for Video Technology  (2023)

\bibitem{Filter-model-based2}
Nosratinia, A.: Embedded post-processing for enhancement of compressed images. In: Proceedings DCC'99 Data Compression Conference (Cat. No. PR00096). pp. 62--71. IEEE (1999)

\bibitem{potlapalli2023promptir}
Potlapalli, V., Zamir, S.W., Khan, S., Khan, F.S.: Promptir: Prompting for all-in-one blind image restoration. arXiv preprint arXiv:2306.13090  (2023)

\bibitem{puigcerver2023sparse-soft-MoE}
Puigcerver, J., Riquelme, C., Mustafa, B., Houlsby, N.: From sparse to soft mixtures of experts. arXiv preprint arXiv:2308.00951  (2023)

\bibitem{rout2024solving-PLSD}
Rout, L., Raoof, N., Daras, G., Caramanis, C., Dimakis, A., Shakkottai, S.: Solving linear inverse problems provably via posterior sampling with latent diffusion models. Advances in Neural Information Processing Systems  \textbf{36} (2024)

\bibitem{saharia2022image-SR3}
Saharia, C., Ho, J., Chan, W., Salimans, T., Fleet, D.J., Norouzi, M.: Image super-resolution via iterative refinement. IEEE Transactions on Pattern Analysis and Machine Intelligence  \textbf{45}(4),  4713--4726 (2022)

\bibitem{schuhmann2022laion-5b}
Schuhmann, C., Beaumont, R., Vencu, R., Gordon, C., Wightman, R., Cherti, M., Coombes, T., Katta, A., Mullis, C., Wortsman, M., et~al.: Laion-5b: An open large-scale dataset for training next generation image-text models. Advances in Neural Information Processing Systems  \textbf{35},  25278--25294 (2022)

\bibitem{shazeer2017outrageously-MoE2}
Shazeer, N., Mirhoseini, A., Maziarz, K., Davis, A., Le, Q., Hinton, G., Dean, J.: Outrageously large neural networks: The sparsely-gated mixture-of-experts layer. arXiv preprint arXiv:1701.06538  (2017)

\bibitem{sheikh2005live1}
Sheikh, H.: Live image quality assessment database release 2. http://live. ece. utexas. edu/research/quality  (2005)

\bibitem{singhal2001modern-cos3}
Singhal, A., et~al.: Modern information retrieval: A brief overview. IEEE Data Eng. Bull.  \textbf{24}(4),  35--43 (2001)

\bibitem{sullivan2012overview-HEVC}
Sullivan, G.J., Ohm, J.R., Han, W.J., Wiegand, T.: Overview of the high efficiency video coding (hevc) standard. IEEE Transactions on circuits and systems for video technology  \textbf{22}(12),  1649--1668 (2012)

\bibitem{vaswani2017attention}
Vaswani, A., Shazeer, N., Parmar, N., Uszkoreit, J., Jones, L., Gomez, A.N., Kaiser, {\L}., Polosukhin, I.: Attention is all you need. Advances in neural information processing systems  \textbf{30} (2017)

\bibitem{wallace1991jpeg}
Wallace, G.K.: The jpeg still picture compression standard. Communications of the ACM  \textbf{34}(4),  30--44 (1991)

\bibitem{wang2023exploring-CLIPIQA}
Wang, J., Chan, K.C., Loy, C.C.: Exploring clip for assessing the look and feel of images. In: Proceedings of the AAAI Conference on Artificial Intelligence. vol.~37, pp. 2555--2563 (2023)

\bibitem{wang2023exploiting-stablesr}
Wang, J., Yue, Z., Zhou, S., Chan, K.C., Loy, C.C.: Exploiting diffusion prior for real-world image super-resolution. arXiv preprint arXiv:2305.07015  (2023)

\bibitem{wang2022jpeg-contrastive}
Wang, X., Fu, X., Zhu, Y., Zha, Z.J.: Jpeg artifacts removal via contrastive representation learning. In: European Conference on Computer Vision. pp. 615--631. Springer (2022)

\bibitem{wang2021realesrgan}
Wang, X., Xie, L., Dong, C., Shan, Y.: Real-esrgan: Training real-world blind super-resolution with pure synthetic data. In: Proceedings of the IEEE/CVF international conference on computer vision. pp. 1905--1914 (2021)

\bibitem{wang2018recovering-spade}
Wang, X., Yu, K., Dong, C., Loy, C.C.: Recovering realistic texture in image super-resolution by deep spatial feature transform. In: Proceedings of the IEEE conference on computer vision and pattern recognition. pp. 606--615 (2018)

\bibitem{wang2022zero-ddnm}
Wang, Y., Yu, J., Zhang, J.: Zero-shot image restoration using denoising diffusion null-space model. arXiv preprint arXiv:2212.00490  (2022)

\bibitem{wang2023dr2}
Wang, Z., Zhang, Z., Zhang, X., Zheng, H., Zhou, M., Zhang, Y., Wang, Y.: Dr2: Diffusion-based robust degradation remover for blind face restoration. In: Proceedings of the IEEE/CVF Conference on Computer Vision and Pattern Recognition. pp. 1704--1713 (2023)

\bibitem{wiegand2003overview-H.264}
Wiegand, T., Sullivan, G.J., Bjontegaard, G., Luthra, A.: Overview of the h. 264/avc video coding standard. IEEE Transactions on circuits and systems for video technology  \textbf{13}(7),  560--576 (2003)

\bibitem{wu2021learned}
Wu, Y., Li, X., Zhang, Z., Jin, X., Chen, Z.: Learned block-based hybrid image compression. IEEE Transactions on Circuits and Systems for Video Technology  \textbf{32}(6),  3978--3990 (2021)

\bibitem{xia2023diffir}
Xia, B., Zhang, Y., Wang, S., Wang, Y., Wu, X., Tian, Y., Yang, W., Van~Gool, L.: Diffir: Efficient diffusion model for image restoration. arXiv preprint arXiv:2303.09472  (2023)

\bibitem{yang2023diffusion-survey-3}
Yang, L., Zhang, Z., Song, Y., Hong, S., Xu, R., Zhao, Y., Zhang, W., Cui, B., Yang, M.H.: Diffusion models: A comprehensive survey of methods and applications. ACM Computing Surveys  \textbf{56}(4),  1--39 (2023)

\bibitem{yang2024ntire-cir}
Yang, R., Timofte, R., Li, B., Li, X., Guo, M., Zhao, S., Zhang, L., Chen, Z., Zhang, D., Arora, Y., et~al.: Ntire 2024 challenge on blind enhancement of compressed image: Methods and results. In: Proceedings of the IEEE/CVF Conference on Computer Vision and Pattern Recognition. pp. 6524--6535 (2024)

\bibitem{yang2022aim}
Yang, R., Timofte, R., Li, X., Zhang, Q., Zhang, L., Liu, F., He, D., Li, F., Zheng, H., Yuan, W., et~al.: Aim 2022 challenge on super-resolution of compressed image and video: Dataset, methods and results. In: European Conference on Computer Vision. pp. 174--202. Springer (2022)

\bibitem{yang2022maniqa}
Yang, S., Wu, T., Shi, S., Lao, S., Gong, Y., Cao, M., Wang, J., Yang, Y.: Maniqa: Multi-dimension attention network for no-reference image quality assessment. In: Proceedings of the IEEE/CVF Conference on Computer Vision and Pattern Recognition. pp. 1191--1200 (2022)

\bibitem{yang2023pixel-PASD}
Yang, T., Ren, P., Xie, X., Zhang, L.: Pixel-aware stable diffusion for realistic image super-resolution and personalized stylization. arXiv preprint arXiv:2308.14469  (2023)

\bibitem{ye2023ip-ipadapter}
Ye, H., Zhang, J., Liu, S., Han, X., Yang, W.: Ip-adapter: Text compatible image prompt adapter for text-to-image diffusion models. arXiv preprint arXiv:2308.06721  (2023)

\bibitem{yee2017medical-BPG}
Yee, D., Soltaninejad, S., Hazarika, D., Mbuyi, G., Barnwal, R., Basu, A.: Medical image compression based on region of interest using better portable graphics (bpg). In: 2017 IEEE international conference on systems, man, and cybernetics (SMC). pp. 216--221. IEEE (2017)

\bibitem{yu2024scaling-SUPIR}
Yu, F., Gu, J., Li, Z., Hu, J., Kong, X., Wang, X., He, J., Qiao, Y., Dong, C.: Scaling up to excellence: Practicing model scaling for photo-realistic image restoration in the wild. arXiv preprint arXiv:2401.13627  (2024)

\bibitem{zamir2022restormer}
Zamir, S.W., Arora, A., Khan, S., Hayat, M., Khan, F.S., Yang, M.H.: Restormer: Efficient transformer for high-resolution image restoration. In: Proceedings of the IEEE/CVF conference on computer vision and pattern recognition. pp. 5728--5739 (2022)

\bibitem{zeyde2012single-classic5}
Zeyde, R., Elad, M., Protter, M.: On single image scale-up using sparse-representations. In: Curves and Surfaces: 7th International Conference, Avignon, France, June 24-30, 2010, Revised Selected Papers 7. pp. 711--730. Springer (2012)

\bibitem{zhang2023adding-Controlnet}
Zhang, L., Rao, A., Agrawala, M.: Adding conditional control to text-to-image diffusion models. In: Proceedings of the IEEE/CVF International Conference on Computer Vision. pp. 3836--3847 (2023)

\bibitem{zhang2018unreasonable-LPIPS}
Zhang, R., Isola, P., Efros, A.A., Shechtman, E., Wang, O.: The unreasonable effectiveness of deep features as a perceptual metric. In: Proceedings of the IEEE conference on computer vision and pattern recognition. pp. 586--595 (2018)

\bibitem{zhao2024uni-adapter}
Zhao, S., Chen, D., Chen, Y.C., Bao, J., Hao, S., Yuan, L., Wong, K.Y.K.: Uni-controlnet: All-in-one control to text-to-image diffusion models. Advances in Neural Information Processing Systems  \textbf{36} (2024)

\bibitem{zhou2022mixture-MoE-latest}
Zhou, Y., Lei, T., Liu, H., Du, N., Huang, Y., Zhao, V., Dai, A.M., Le, Q.V., Laudon, J., et~al.: Mixture-of-experts with expert choice routing. Advances in Neural Information Processing Systems  \textbf{35},  7103--7114 (2022)

\bibitem{zhu2023prompt-prompt-latest2}
Zhu, B., Niu, Y., Han, Y., Wu, Y., Zhang, H.: Prompt-aligned gradient for prompt tuning. In: Proceedings of the IEEE/CVF International Conference on Computer Vision. pp. 15659--15669 (2023)

\bibitem{zhu2023designing-decoder-compensator}
Zhu, Z., Feng, X., Chen, D., Bao, J., Wang, L., Chen, Y., Yuan, L., Hua, G.: Designing a better asymmetric vqgan for stablediffusion. arXiv preprint arXiv:2306.04632  (2023)

\end{thebibliography}

\appendix
\section*{Appendix}
In this \textbf{Appendix}, we first illustrate more details of MoE-Prompt in Sec.~\ref{Appendix:MoE Prompts}. Then, we provide more quantitative results in Sec.~\ref{Appendix:more quantitative studies}. Finally, we show more visual comparisons in Sec.~\ref{Appendix:more visual Results}.

\section{More Details of MoE-Prompt}
\label{Appendix:MoE Prompts}
In this section, we provide additional details of MoE-Prompt Module.
\begin{figure}[htp]
	\centering
	\includegraphics[width=1\linewidth]{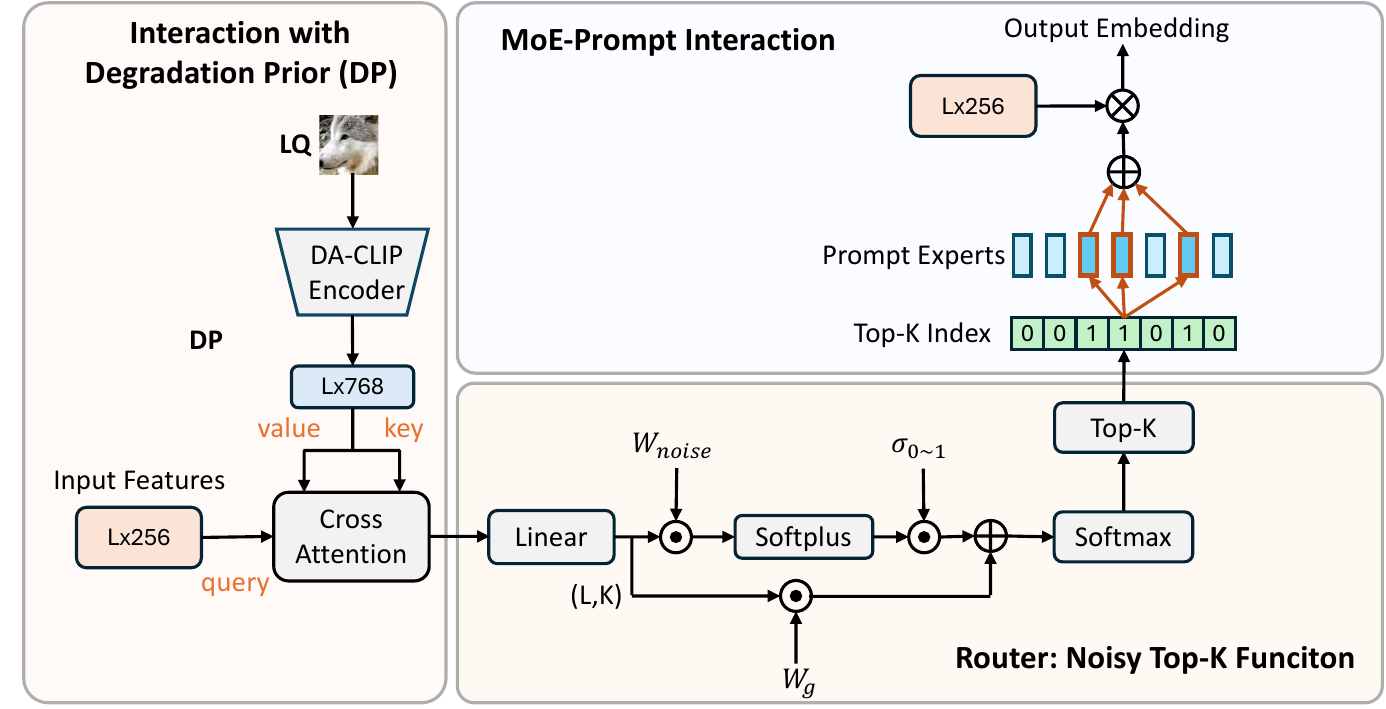}
	\caption{More details about working pipeline in MoE-Prompt Module. Input features interact with degradation prior through Cross Attention~\cite{vaswani2017attention}. Then router uses Noisy Top-K function to choose a combination of K prompts, which are first aggregated through summation and then multiplied to input features. }
\label{fig:Supp}
\end{figure}

As depicted in Fig.~\ref{fig:Supp}, we use pre-trained image controller offered by DACLIP~\cite{luo2023controlling-DACLIP} to extract distortion information from low quality image. It is worth noting that DACLIP, having been trained on various tasks, possesses robustness across multiple tasks, including the ability to extract features specific to individuals. Consequently, we did not fine-tune this model on our CIR dataset. This degradation prior (DP) interacts with input features through Cross Attention~\cite{vaswani2017attention}. Subsequently, we select a combination of K prompts using the Noisy Top-K function~\cite{shazeer2017outrageously-MoE2} mentioned in the main text. Here $W_{noise}$ and $W_{g}$ are learnable parameters. These K prompts are first aggregated through summation, then followed by a matrix multiplication with the input features, ultimately yielding the output embedding. 

\begin{figure}[t]
	\centering
	\includegraphics[width=1\linewidth]{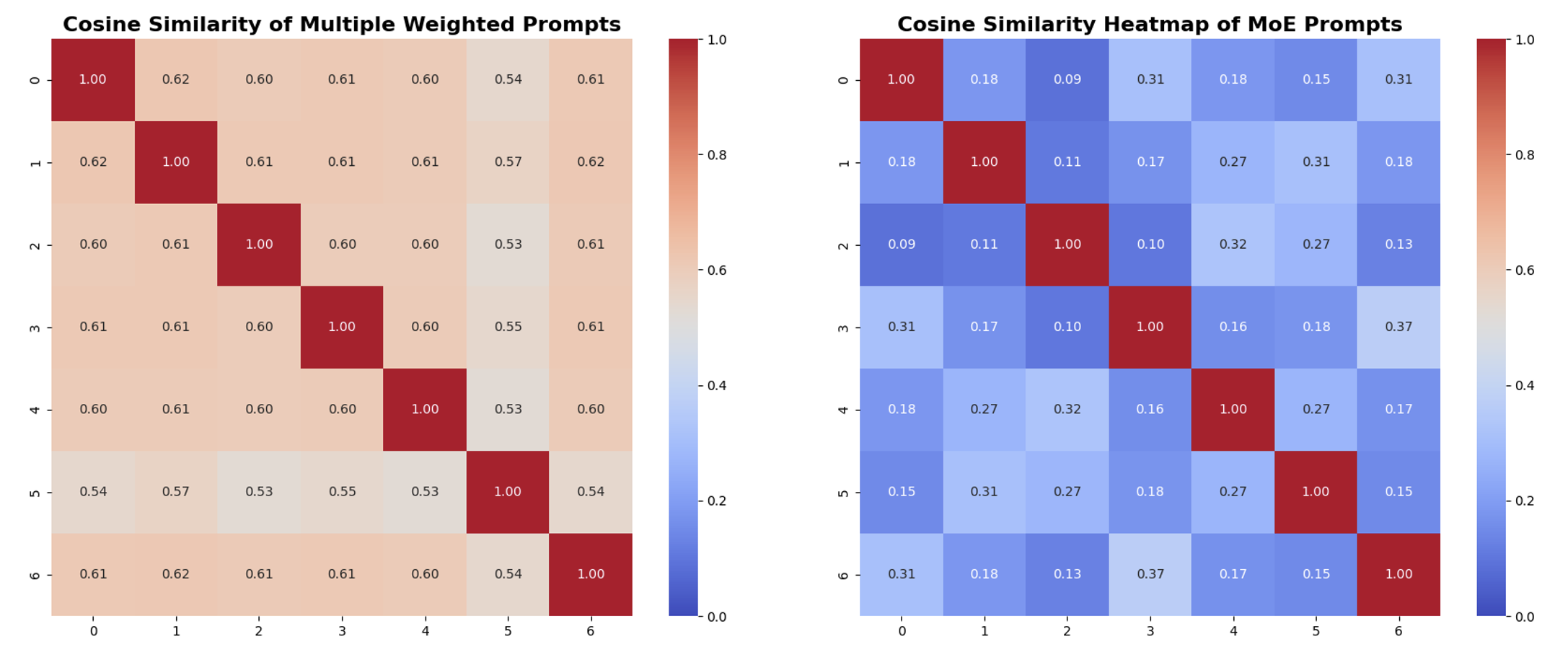}
	\caption{ A comparison of cosine similarity heatmaps for two different designs of prompts: Multiple Weighted Prompts and MoE (Mixture of Experts) prompts. Each heatmap visualizes the pairwise cosine similarities among seven prompts, with color intensity reflecting the degree of similarity.}
	\label{fig:Supp2}
\end{figure}

In this section, we further demonstrate the effectiveness of our proposed Mixture of Experts (MoE) Prompts. In particular, compared to the Single Prompt~\cite{ma2023prores} or Multiple Weighted Prompts approaches~\cite{potlapalli2023promptir,ai2023multimodal-mperceiver,li2023prompt-PIP,luo2023controlling-DACLIP}, our MoE-based scheduling mechanism enables prompts to independently learn more informative and distinct features of different compression distortions. Accordingly, we measure the cosine similarity between different prompts as a means to represent whether the prompts have learned independent features~\cite{aggarwal2001surprising-cos1,dubey2016cosine-cos2,singhal2001modern-cos3}. In this context, a higher cosine similarity indicates that the prompts have learned more uniform knowledge, while a lower cosine similarity suggests that each prompt has acquired more independent information, corresponding to different types of distortion. The comparison results with multiple weighted prompts are illustrated in Fig.~\ref{fig:Supp2}. It can be observed that 7 prompts in the multiple weighted prompts approach are quite similar, indicating that they have learned more homogenized distortion information, leading to low utilization among the prompts. In contrast, the feature similarity among our proposed MoE-Prompt is significantly lower, suggesting that each prompt is more independent and better equipped to handle various distortions, thereby maximizing the utilization of the prompts.

\section{More Quantitative Results}
\label{Appendix:more quantitative studies}
\subsection{The effect of VAE fine-tuning}
Here, we present the PSNR values in Table~\ref{table:review1} of the reconstructed images in three designated settings: using only pre-trained VAE, training a VAE from scratch, and employing the Decoder Compensator (our method). From Table 1, it is evident that our method significantly enhances fidelity in terms of objective metrics. Training the VAE from scratch yields inferior results, primarily due to the substantial data requirement for achieving a well-performing VAE.

\begin{table}[h!]
\centering
\caption{Different VAE training settings (The rightmost is our method). Results are tested on LIVE1~\cite{sheikh2005live1}.}
\resizebox{\linewidth}{!}{
\begin{tabular}{c|c|c|c}
\Xhline{2pt}
VAE Settings                    & Not fine-tune or train VAE & Train VAE from scrarch & Fine-tune  VAE decoder (Ours) \\ \hline
PSNR/SSIM (Average on 21 tasks) & 22.70/0.646               & 25.12/0.705            & 29.03/0.812       \\    
\Xhline{2pt}
\end{tabular}}
\label{table:review1}
\end{table}

\subsection{More detailed comparison with other methods}
In this section, we provided the detailed experimental results for each compression configuration. Here, considering the fast development of image restoration~\cite{kawar2022denoising-ddrm,jin2021dc,li2023diffusion-survey,chung2022diffusion-dps}, we compare our methods with several typical IR methods, including one GAN-based methods RealESRGAN~\cite{wang2021realesrgan}, one transformer-based method PromptIR~\cite{potlapalli2023promptir}, four Diffusion-based methods StableSR~\cite{wang2023exploiting-stablesr}, DiffBIR~\cite{lin2023diffbir}, PASD~\cite{yang2023pixel-PASD} and SUPIR~\cite{yu2024scaling-SUPIR}. All methods are reproduced on our proposed CIR dataset. However, it is worth noting that, as SUPIR does not provide the official training code, we directly utilize the pre-trained model offered by the code repositories of SUPIR for sampling purposes. To further assess the subjective quality of the generated images, we also expand our perceptual metrics to include two no-reference (NR) metrics: ClipIQA~\cite{wang2023exploring-CLIPIQA} and ManIQA~\cite{yang2022maniqa}. Here, we use IQA-PyTorch~\cite{pyiqa} to implement these metrics with model card 'clipiqa+' and 'maniqa-pipal' respectively. We conduct comprehensive tests for each codec across three different levels of distortion. Results are presented from Table~\ref{table:JPEG} to Table~\ref{table:HIFIC}. It is observed that our proposed MoE-DiffIR not only significantly outperforms other diffusion-based models in metrics with Ground Truth, such as PSNR, SSIM, LPIPS, and FID, but also demonstrates competitive strength in the no-reference (NR) metrics: ClipIQA and ManIQA.

\begin{table*}[t]
\centering
\caption{Quantitative comparison for compressed image restoration on JPEG~\cite{wallace1991jpeg} Codec with three distortion levels. Results are tested on with different metrics in terms of  PSNR$\uparrow$, SSIM$\uparrow$, LPIPS$\downarrow$, FID$\downarrow$, ClipIQA$\uparrow$ and ManIQA$\uparrow$. Red and blue colors represent the best and second best performance, respectively.(Suggested to zoom in for better visualization.) \textit{All comparison methods are reproduced on our constructed CIR datasets except for SUPIR~\cite{yu2024scaling-SUPIR}.}}
\setlength{\tabcolsep}{2pt}
\resizebox{\textwidth}{!}{
\begin{tabular}{c|c|cccccccccccccccccc}
\Xhline{2pt}
                           &                           & \multicolumn{18}{c}{JPEG}                                                                                                                                                                                                                                                                                                                                                                                                                                                                                                                                                                                                        \\ \cline{3-20} 
                           &                           & \multicolumn{6}{c|}{Q=10}                                                                                                                                                                                        & \multicolumn{6}{c|}{Q=15}                                                                                                                                                                                        & \multicolumn{6}{c}{Q=20}                                                                                                                                                                   \\
\end{tabular}
}
\label{table:JPEG}
\end{table*}

\begin{table*}[t]
\centering
\caption{Quantitative comparison for compressed image restoration on VVC~\cite{bross2021overview-VVC} Codec with three distortion levels. Red and blue colors represent the best and second best performance, respectively.(Suggested to zoom in for better visualization.) \textit{All comparison methods are reproduced on our constructed CIR datasets except for SUPIR~\cite{yu2024scaling-SUPIR}.}}
\setlength{\tabcolsep}{2pt}
\resizebox{\textwidth}{!}{
\begin{tabular}{c|c|cccccccccccccccccc}
\Xhline{2pt}
                           &                           & \multicolumn{18}{c}{VVC}                                                                                                                                                                                                                                                                                                                                                                                                                                                                                                                                                                                                         \\ \cline{3-20} 
                           &                           & \multicolumn{6}{c|}{Q=47}                                                                                                                                                                                        & \multicolumn{6}{c|}{Q=42}                                                                                                                                                                                        & \multicolumn{6}{c}{Q=37}                                                                                                                                                                   \\
\end{tabular}
}
\label{table:VVC}
\end{table*}
\setlength{\intextsep}{5pt}

\begin{table*}[t]
\centering
\caption{Quantitative comparison for compressed image restoration on HEVC~\cite{sullivan2012overview-HEVC} Codec with three distortion levels. Red and blue colors represent the best and second best performance, respectively.(Suggested to zoom in for better visualization.) \textit{All comparison methods are reproduced on our constructed CIR datasets except for SUPIR~\cite{yu2024scaling-SUPIR}.}}
\setlength{\tabcolsep}{2pt}
\resizebox{\textwidth}{!}{
\begin{tabular}{c|c|cccccccccccccccccc}
\Xhline{2pt}
                           &                           & \multicolumn{18}{c}{HEVC}                                                                                                                                                                                                                                                                                                                                                                                                                                                                                                                                                                                                         \\ \cline{3-20} 
                           &                           & \multicolumn{6}{c|}{Q=47}                                                                                                                                                                                        & \multicolumn{6}{c|}{Q=42}                                                                                                                                                                                        & \multicolumn{6}{c}{Q=37}                                                                                                                                                                    \\
\end{tabular}
}
\label{table:HEVC}
\end{table*}
\setlength{\intextsep}{5pt}

\begin{table*}[t]
\centering
\caption{Quantitative comparison for compressed image restoration on WebP~\cite{ginesu2012objective-WEBP}
Codec with three distortion levels. Red and blue colors represent the best and second best performance, respectively.(Suggested to zoom in for better visualization.) \textit{All comparison methods are reproduced on our constructed CIR datasets except for SUPIR~\cite{yu2024scaling-SUPIR}.}}
\setlength{\tabcolsep}{2pt}
\resizebox{\textwidth}{!}{
\begin{tabular}{c|c|cccccccccccccccccc}
\Xhline{2pt}
                           &                           & \multicolumn{18}{c}{WebP}                                                                                                                                                                                                                                                                                                                                                                                                                                                                                                                                                                                                         \\ \cline{3-20} 
                           &                           & \multicolumn{6}{c|}{Q=1}                                                                                                                                                                                        & \multicolumn{6}{c|}{Q=5}                                                                                                                                                                                        & \multicolumn{6}{c}{Q=10}                                                                                                                                                                    \\
\end{tabular}
}
\label{table:WebP}
\end{table*}
\setlength{\intextsep}{5pt}

\begin{table*}[t]
\centering
\caption{Quantitative comparison for compressed image restoration on $C_{PSNR}$~\cite{cheng2020learned}
Codec with three distortion levels. Red and blue colors represent the best and second best performance, respectively.(Suggested to zoom in for better visualization.) \textit{All comparison methods are reproduced on our constructed CIR datasets except for SUPIR~\cite{yu2024scaling-SUPIR}.}}
\setlength{\tabcolsep}{2pt}
\resizebox{\textwidth}{!}{
\begin{tabular}{c|c|cccccccccccccccccc}
\Xhline{2pt}
                           &                           & \multicolumn{18}{c}{$C_{PSNR}$}                                                                                                                                                                                                                                                                                                                                                                                                                                                                                                                                                                                                         \\ \cline{3-20} 
                           &                           & \multicolumn{6}{c|}{Q=1}                                                                                                                                                                                        & \multicolumn{6}{c|}{Q=2}                                                                                                                                                                                        & \multicolumn{6}{c}{Q=3}                                                                                                                                                                    \\
\end{tabular}
}
\label{table:PSNR}
\end{table*}
\setlength{\intextsep}{5pt}

\begin{table*}[t]
\centering
\caption{Quantitative comparison for compressed image restoration on $C_{SSIM}$~\cite{cheng2020learned}
Codec with three distortion levels. Red and blue colors represent the best and second best performance, respectively.(Suggested to zoom in for better visualization.) \textit{All comparison methods are reproduced on our constructed CIR datasets except for SUPIR~\cite{yu2024scaling-SUPIR}.}}
\setlength{\tabcolsep}{2pt}
\resizebox{\textwidth}{!}{
\begin{tabular}{c|c|cccccccccccccccccc}
\Xhline{2pt}
                           &                           & \multicolumn{18}{c}{$C_{SSIM}$}                                                                                                                                                                                                                                                                                                                                                                                                                                                                                                                                                                                                      \\ \cline{3-20} 
                           &                           & \multicolumn{6}{c|}{Q=1}                                                                                                                                                                                        & \multicolumn{6}{c|}{Q=2}                                                                                                                                                                                        & \multicolumn{6}{c}{Q=3}                                                                                                                                                                    \\
\end{tabular}

}
\label{table:SSIM}
\end{table*}
\setlength{\intextsep}{5pt}

\begin{table*}[t]
\centering
\caption{Quantitative comparison for compressed image restoration on HIFIC~\cite{mentzer2020high-HIFIC}
Codec with three distortion levels. Red and blue colors represent the best and second best performance, respectively.(Suggested to zoom in for better visualization.) \textit{All comparison methods are reproduced on our constructed CIR datasets except for SUPIR~\cite{yu2024scaling-SUPIR}.}}
\setlength{\tabcolsep}{2pt}
\resizebox{\textwidth}{!}{
\begin{tabular}{c|c|cccccccccccccccccc}
\Xhline{2pt}
                           &                           & \multicolumn{18}{c}{HIFIC}                                                                                                                                                                                                                                                                                                                                                                                                                                                                                                                                                                                                    \\ \cline{3-20} 
                           &                           & \multicolumn{6}{c|}{Mode = 'Low'}                                                                                                                                                                                & \multicolumn{6}{c|}{Mode = 'Med'}                                                                                                                                                                               & \multicolumn{6}{c}{Mode = 'High'}                                                                                                                                                          \\
\end{tabular}
}
\label{table:HIFIC}
\end{table*}
\setlength{\intextsep}{5pt}

\section{More Visual Results}
\label{Appendix:more visual Results}
\begin{figure}[t]
	\centering
	\includegraphics[width=1\linewidth]{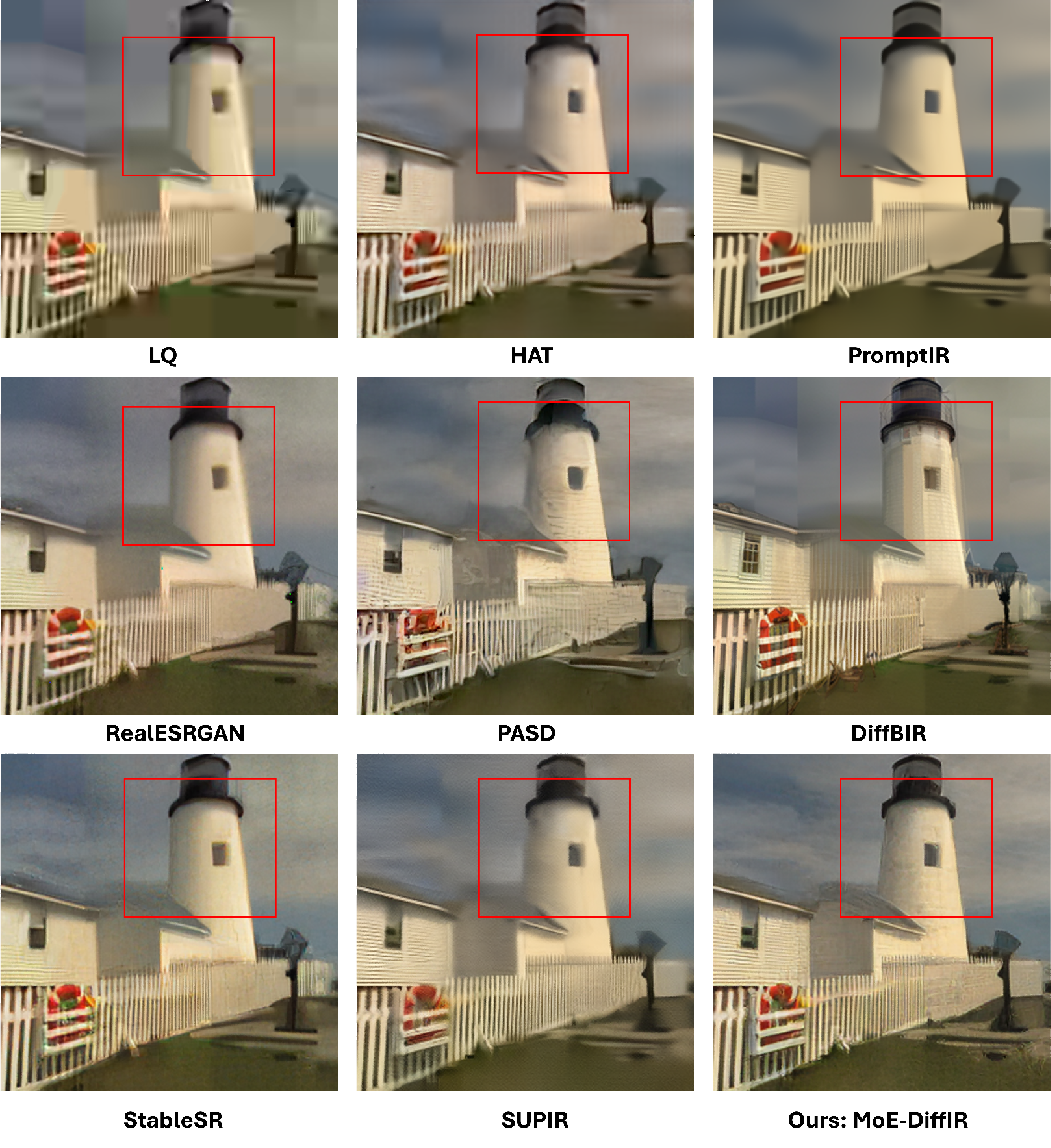}
	\caption{Visual comparisons of our MoE-DiffIR with other SOTA models on codec VTM(QP=47).}
	\label{fig:viusal2}
\end{figure}

\begin{figure}[t]
	\centering
	\includegraphics[width=1\linewidth]{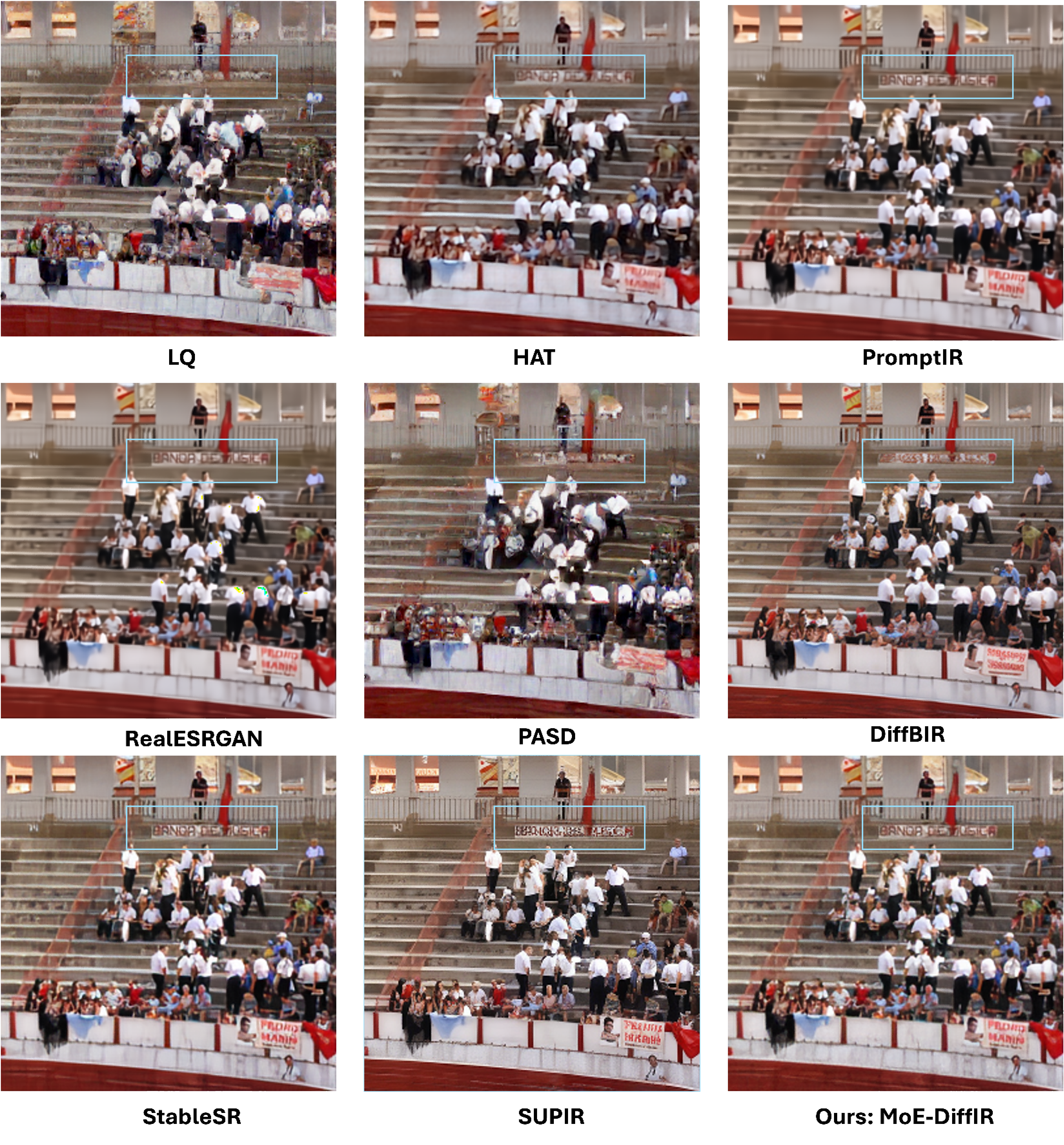}
	\caption{Visual comparisons of our MoE-DiffIR with other SOTA models on codec $C_{PNSR}$(Q=3).}
	\label{fig:viusal1}
\end{figure}

\begin{figure}[t]
	\centering
	\includegraphics[width=1\linewidth]{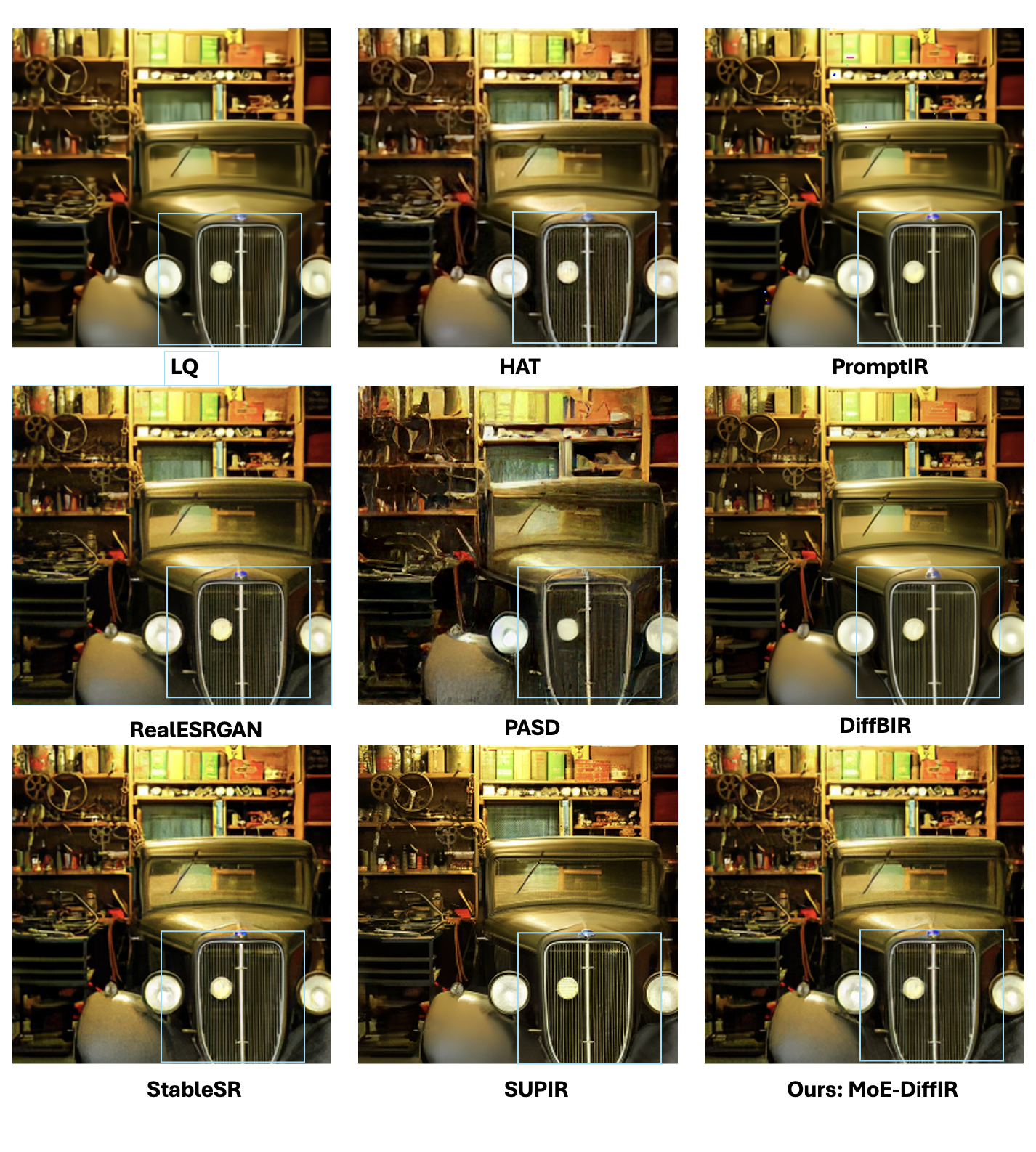}
	\caption{Visual comparisons of our MoE-DiffIR with other SOTA models on codec HM (QP=37).}
	\label{fig:viusal3}
\end{figure}

We provide additional visual comparisons for MoE-DiffIR on Compressed Image Restoration(CIR) tasks. The results are shown in Fig.~\ref{fig:viusal2}, Fig.~\ref{fig:viusal1} and Fig.~\ref{fig:viusal3}. We conducted tests from low bitrates (HM, QP=37) to high bitrates (VTM, QP=47). From the visual results, we can observe that at higher bitrates, our model excels in text generation compared to other diffusion-model-based restoration methods. Models like SUPIR~\cite{yu2024scaling-SUPIR}, DiffBIR~\cite{lin2023diffbir}, and PASD~\cite{yang2023pixel-PASD} may generate incorrect texture information. However, our MoE-DiffIR, benefiting from further fine-tuning of the decoder using MoE-Prompt, produces details that are more faithful to the original image. Additionally, at lower bitrates, as shown in Fig.~\ref{fig
}, models such as DiffBIR~\cite{lin2023diffbir} and SUPIR~\cite{yu2024scaling-SUPIR} fail to eliminate background noise interference within the red boxes, whereas our model achieves noticeable restoration effects. This demonstrates the robustness of our MoE-Prompt for universal tasks.

\end{document}